\documentclass[preprint,12pt,authoryear]{elsarticle}
\usepackage{microtype,setspace}
\usepackage[pdftex,colorlinks=true,unicode]{hyperref}
\usepackage{amssymb,booktabs,subfig,xcolor,paralist, makecell, longtable, bm, orcidlink, dsfont}
\usepackage[top=1.5in, bottom=1.5in, left=1.2in, right=1.2in]{geometry}
\usepackage{natbib}
\usepackage{url} 
\definecolor{darkblue}{rgb}{0,0,.6}
\hypersetup{citecolor=darkblue,linkcolor=darkblue,urlcolor=darkblue}
\usepackage{amsmath}
\usepackage{amsfonts}
\usepackage{float}
\usepackage{epsfig}
\usepackage{graphics}
\usepackage[T1]{fontenc}
\usepackage{lmodern}
\usepackage{graphicx}
\usepackage{lscape}
\usepackage{booktabs}
\usepackage{rotating}
\usepackage[normalem]{ulem}
\usepackage{soul}

\usepackage{orcidlink}
\allowdisplaybreaks[4]
\DeclareMathOperator*{\argmin}{arg\,min}

\newcommand{\X}{\mathcal{X}}

\newsavebox\CBox

\usepackage[font=small]{caption}

\graphicspath{{plots/}}

\newcommand{\Rlogo}{\protect\includegraphics[height=1.8ex,keepaspectratio]{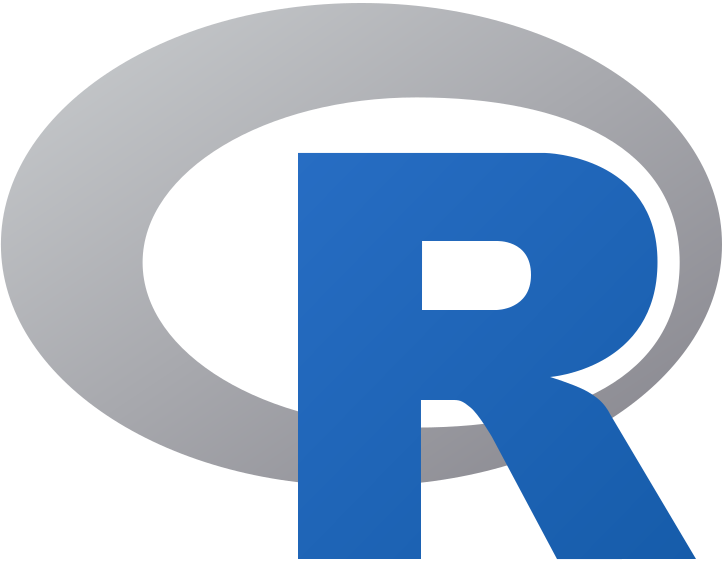}}

\usepackage{amsthm,thmtools}
\usepackage{mathrsfs}

\def\spacingset#1{\renewcommand{\baselinestretch}%
{#1}\small\normalsize} \spacingset{1}

\newcommand{\gammaest}{\widehat{\bm{\gamma}}}
\newcommand{\gammanosparse}{\widehat{\bm{\gamma}}(0,\lambda_{2})}
\newcommand{\gammatrue}{\bm{\gamma}_{\star}}

\newcommand{\gammanosparseG}{\widehat{\bm{\gamma}}_{g}(0,\lambda_{2})}
\newcommand{\gammaestG}{\widehat{\bm{\gamma}}_{g}}

\declaretheorem{theorem}

\makeatletter
\def\th@newremark{\th@remark\thm@headfont{\bfseries}}
\makeatletter
\theoremstyle{newremark}

\newtheorem{assumption}{Assumption}
\declaretheoremstyle[
  spaceabove=6pt, spacebelow=6pt,
  headfont=\bfseries,
  notefont=\mdseries, notebraces={(}{)},
bodyfont=\normalfont,
  postheadspace=0.5em,
]{mystyle}

\begin{document}
\begin{frontmatter}
\title{Interpretable additive model for analyzing high-dimensional functional time series}

\author[1]{Haixu Wang\corref{cor1}\orcidlink{0009-0001-6315-5187}}
\ead{haixu.wang@ucalgary.ca}
\cortext[cor1]{Corresponding author}
\affiliation[1]{organization={Department of Mathematics and Statistics},
addressline={University of Calgary},
city={Calgary},
state={Alberta},
postcode={T2N 1N4},
country={Canada}}

\author[2]{Tianyu Guan}
\ead{tguan@yorku.ca}
\affiliation[2]{organization={Department of Mathematics and Statistics},
addressline={York University},
city={Toronto},
state={Ontario},
postcode={M3J 1P3},
country={Canada}}

\author[3]{Han Lin Shang\orcidlink{0000-0003-1769-6430}}
\ead{hanlin.shang@mq.edu.au}
\affiliation[3]{organization={Department of Actuarial Studies and Business Analytics},
addressline={Macquarie University},
city={Sydney},
state={New South Wales},
postcode={2109},
country={Australia}}


\begin{abstract}
High-dimensional functional time series offer a powerful framework for extending functional time series analysis to settings with multiple simultaneous dimensions, capturing both temporal dynamics and cross-sectional dependencies. We propose a novel, interpretable additive model tailored for such data, designed to deliver both high predictive accuracy and clear interpretability. The model features bivariate coefficient surfaces to represent relationships across panel dimensions, with sparsity introduced via penalized smoothing and group bridge regression. This enables simultaneous estimation of the surfaces and identification of significant inter-dimensional effects. Through Monte Carlo simulations and an empirical application to Japanese subnational age-specific mortality rates, we demonstrate the proposed model’s superior forecasting performance and interpretability compared to existing functional time series approaches.
\end{abstract}

\begin{keyword}
group bridge regression; high-dimensional functional time series; interpretable model; triangulation; subnational mortality rates
\end{keyword} 

\end{frontmatter}

\section{Introduction}

\begin{figure}[!htb]
\centering
\includegraphics[width=.84\textwidth]{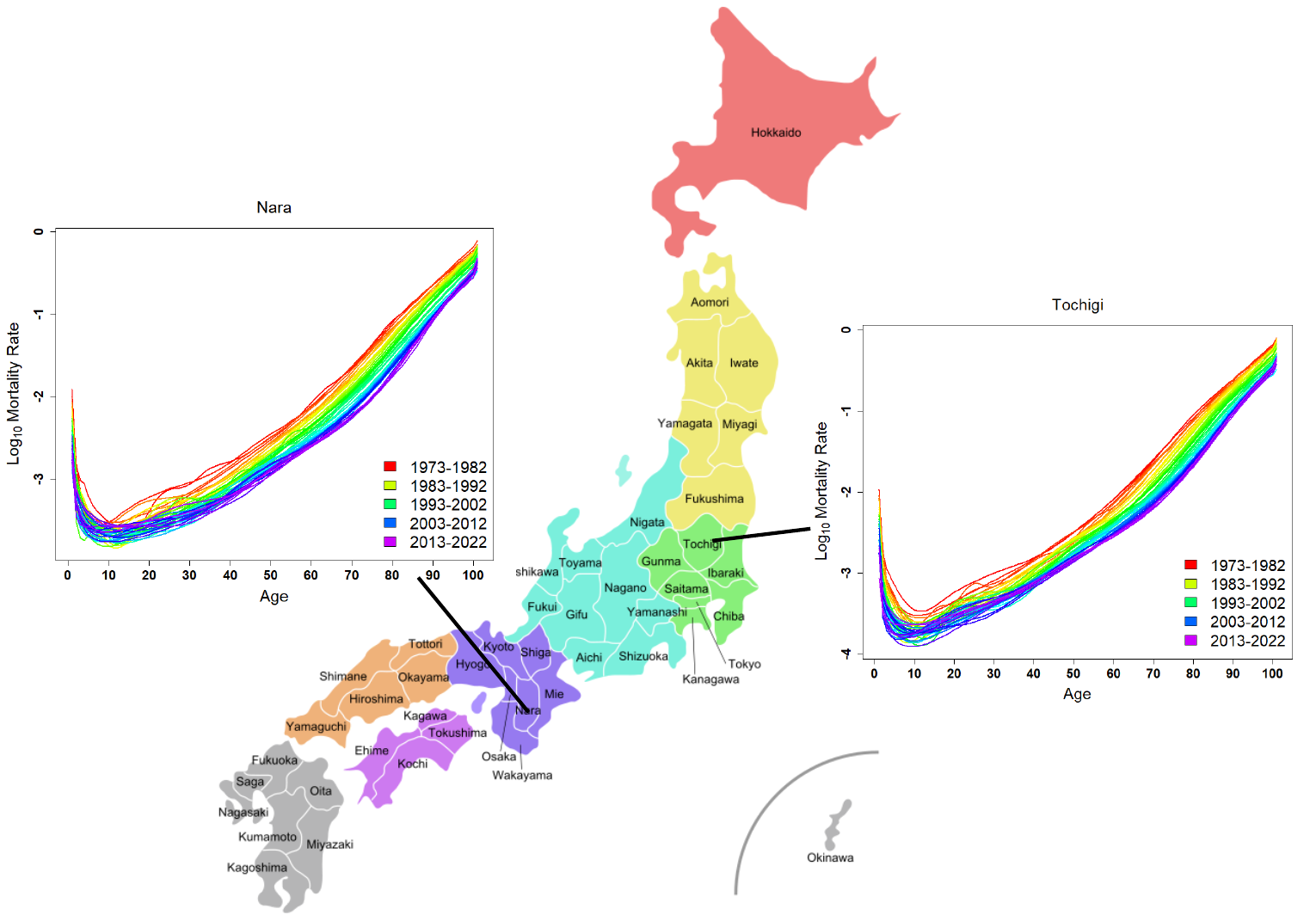}
\caption{Smoothed \texorpdfstring{$\log_{10}$}{log10} mortality rate curves from the year 1973 to 2022 for two prefectures \texorpdfstring{\textbf{Nara}}{Nara} and \texorpdfstring{\textbf{Tochigi}}{Tochigi} in Japan. The rainbow color represents the year of the mortality rate curve, ranging from red (earliest) to purple (most current).}\label{fig:data_intro}
\end{figure}

High-dimensional functional time series (HDFTS) have gained increasing attention because of their ability to capture complex temporal dynamics and cross-sectional dependencies. A representative example is age-specific mortality rates observed in Japan's 47 prefectures over several decades. Figure~\ref{fig:data_intro} presents smoothed $\mathrm{log}_{10}$ mortality rate curves from 1973 to 2022 for two randomly selected prefectures. In this study, our objective is to explore how mortality trends in one prefecture can be influenced by historical patterns in neighboring regions. To this end, we propose a novel interpretable additive model that not only delivers strong predictive performance, but also identifies the specific prefectures and age ranges that significantly contribute to mortality forecasts. These inter-temporal and inter-regional effects are captured through bivariate coefficient surfaces, offering an interpretable representation of additive influences across time and space.

In recent years, there has been significant progress in the analysis of HDFTS. For example, \citet{ZD23} developed Gaussian and multiplier bootstrap approximations for the sums of HDFTS, which enable the construction of joint simultaneous confidence bands for mean functions and support hypothesis testing to assess parallel behavior across the cross-sectional dimension. \citet{HNT231} explored the factor representation of HDFTS, establishing key conditions on the eigenvalues of the covariance operator necessary for the existence and uniqueness of a factor model.

Several factor modeling approaches have been proposed. \citet{GSY19} introduced a two-stage method that applies truncated principal component analysis followed by a scalar factor model on the panel of scores. \citet{HNT232} proposed a functional factor model featuring functional factor loadings and a vector of real-valued factors, while \citet{GQW22} introduced a complementary approach with real-valued factor loadings and functional factors. \citet{LLS+24} unified these models under a single framework that accommodates both types of structures. Under the unified factor model, \cite{LLP25} proposed an easy-to-implement criterion to consistently select the number of common stochastic trends and further discussed model estimation when nonstationary factors are co-integrated.

Beyond factor modeling, \citet{TSY22} and \cite{LSS25} addressed the clustering of age-specific subnational mortality rates, which is an important application of HDFTS. \citet{LLS24} developed hypothesis tests for the detection, estimation and grouping of change points using an information criterion suited for HDFTS. For forecasting HDFTS, refer to \citet{JSS24a}, \citet{CFQ+25} and \cite{TSY+25}. Furthermore, \cite{guo2023consistency} introduced a vector functional autoregressive model under high-dimensional settings, considering both temporal dependencies within the panel and panel-specific dependencies. In addition, \cite{xu2024modeling} proposed using partial covariance operators for functional autoregressive models.

Meanwhile, important research directions in functional regression and variable selection continue to attract substantial attention within the statistical community. A central objective is to develop regression frameworks that link covariates (functional or scalar) to responses that may also be functional or scalar. Beyond identifying significant predictors, variable selection in this context involves determining where within the domain of a functional covariate its influence is most pronounced. This localization of effect is a distinctive challenge in functional data analysis and functional regression \citep[see, e.g.,][]{Morris15, RGS+17}.

Variable selection in functional regression presents a two-fold challenge: identifying significant predictors (global selection) and determining the specific regions within those predictors that contribute meaningfully to the response (local selection). The global selection problem involves selecting a subset of relevant covariates (functional or scalar) from a larger pool. For example, \citet{PFLR} proposed a unified framework for selecting scalar and functional predictors, while \citet{funvariableselection} provided a comprehensive review of variable selection methods in functional regression. Additional discussions on global selection can be found in \cite{partialLM}, \cite{Lian2013}, \cite{huangetal2016}, and \cite{maetal2019}, while \citet{fan2015functional} offers theoretical insights into variable selection in functional linear models.

In contrast, local variable selection focuses on identifying influential subregions within the support of a functional covariate. This problem is inherently more complex because it involves determining where, rather than which, predictors have significant effects. A simplified formulation considers regions where the functional regression coefficients are zero, whether in intervals, patches, or other subsets of the domain. The theoretical foundations of this approach are discussed in \citet{huanghorowitzwei2010}, while \citet{MS10} and \citet{KPS16} introduce the concept of ``points of impact," where a predictor's influence is localized. \citet{interpretableFLR} formally connected interpretability with the local sparsity of regression coefficients, and \citet{globalvslocal} presented global versus local sparsity structures. Furthermore, \citet{fscad} explored the use of the SCAD penalty as an alternative to standard $L_1$ regularization to achieve sparsity in functional regression.

In this work, our objective is to develop functional regression models tailored for HDFTS, with an emphasis on achieving strong predictive performance and interpretability. Although functional data analysis has seen significant advances in both prediction and variable selection, existing approaches face limitations when applied directly to HDFTS.

Most current methods are location-specific; that is, models are built using data from a particular region and are used to make predictions for that same region. Such approaches fall short in the context of HDFTS, where it is more appropriate to leverage historical information from all regions to predict the functional response for a given location. Furthermore, the relationships between regions are often heterogeneous, meaning that a single global model cannot adequately capture the region-specific dynamics.

To address this, we propose a framework (\textcolor{blue}{see model~\eqref{eq:1}}) in which a unique region-specific model is constructed for each location. Each model is informed by historical data from all regions (including the target region) and may incorporate long-term dependencies across time. This approach offers greater flexibility and improved predictive accuracy. However, it comes with trade-offs in terms of interpretability and computational efficiency, which we aim to manage through careful model design and regularization.

We aim to develop a model for HDFTS that balances predictive performance with interpretability. Achieving this requires a focus on model interpretability, addressing a key limitation in existing approaches. Typically, random functions in the panel data are reduced to a set of linear projections, and prediction models are based on these coefficients. However, this method obscures the direct relationship between the predictors and the response, as the connection between the linear projections of both is not easily interpretable. To overcome this, we propose maintaining the original functional forms of both the predictors and the response, instead of relying on their projections. Our novel additive model is designed to ensure accurate forecasting and clear interpretability, making the relationships within the HDFTS transparent and accessible.

In introducing a novel and interpretable model for predicting HDFTS, designed to enhance both predictive performance and interpretability, we make four key contributions. First, we employ an additive model that incorporates information from all regions to predict the functional responses of a specific location in the panel. Unlike projection-based methods \citep[e.g.,][]{guo2016spline, xu2024modeling} that model dependencies through the covariance of basis expansion scores, we account for cross-dimensional dependencies by using multiple functional data directly rather than through their projections. This unveils a more direct relationship, which cannot be explained through its surrogate based on projections of functional data. HDFTS are often associated with both physical locations and time (e.g., functional time series across various locations). Our model is especially suited to this scenario as it captures both temporal dependencies within the data and cross-sectional dependencies across regions.

Second, we do not reduce the HDFTS to a set of linear projections. Instead, the regression model is estimated in its original form, preserving the direct relationship between predictors and the response. Interpretability is achieved through the regression coefficients, which are represented as bivariate functions or surfaces. This approach is more straightforward and eliminates the need for manual transformations of the HDFTS.

Although our model is complex and interpretable, it introduces potential challenges of overfitting and computational complexity. In the context of HDFTS, overfitting can arise when the number of training data pairs is smaller than the number of predictors. In HDFTS across different locations, the number of regions often exceeds the number of time points. In such cases, overfitting occurs, making the predictions unreliable. Thus our third contribution is in addressing the global selection problem by employing penalization techniques to select a subset of predictors for model construction, addressing the global selection problem. This leads to a better interpretation of the model, highlighting the relationships between different regions and distinguishing significant from non-significant predictors.

Finally, our fourth contribution is enhancing model interpretability through local variable selection, embedded within the global process. In selecting the significant influences, we also identify the regions of the predictors that most significantly contribute to the response. This unique feature, which is not available in existing approaches, allows for a deeper understanding of how the predictors from different regions influence the response.

Our model strikes an optimal balance between computational efficiency and predictive performance. In comparison to nonlinear machine learning methods, such as those outlined by \cite{NOP}, our approach delivers competitive predictive accuracy without the computational burden or risk of over-parameterization typically associated with these methods. While machine learning-based approaches may offer superior predictive performance, they often demand significant computational time and resources. Our model, however, achieves equivalent predictive performance while maintaining the computational efficiency characteristic of traditional statistical models. Through a series of Monte Carlo simulations, we demonstrate the model’s consistency in estimation and its robust predictive capabilities. Additionally, we apply the proposed model to Japanese age-specific mortality rates, uncovering regional interactions, age-specific effects, and temporal lag effects. This application highlights the model’s ability to provide valuable insights into mortality trends across regions, underscoring its practical utility in real-world applications.

The remainder of this paper is organized as follows. Section~\ref{sec:2} provides a detailed specification of the model and the estimation approach. Theoretical results are presented in Section~\ref{sec:theo}. Section~\ref{sec:simu} discusses the Monte Carlo simulation results, which illustrate the numerical performance of the model. In Section~\ref{sec:4}, we apply our model to subnational age-specific mortality rates in Japan, showcasing how the model captures regional and age-dependent interactions. Finally, Section~\ref{sec:5} concludes the paper, summarizing the findings and suggesting directions for future research.

\section{Model specification and estimation}\label{sec:2}

We begin by introducing the HDFTS and our prediction model. Let $\{\X_{ts}(u)\}_{t=1, s = 1}^{t = n, s = S}$ be a square-integrable function over some interval $u\in \mathcal{I} \subset \mathbb{R}$, where $\mathcal{I}$ denotes a compact function support, which is a subset of the real-valued space, $t$ is the discrete index of time, and $s$ is the index of region (reflecting the dimension of HDFTS). The forecast model for the $s$\textsuperscript{th} region takes the following additive form:
\begin{equation}
\X_{ts}(v) = \sum_{g=1}^{S}\int_{u\in \mathcal{I}}\beta_{sg}(u,v)\X_{t-\delta,g}(u)du + \epsilon_{ts}(v),\label{eq:1}
\end{equation}
where $\delta$ represents the time lag between a pair of observations in the time series, and $\epsilon_{ts}(v)$ denotes error term. It is common practice to set $\delta = 1$ for the time lag in predictions, and we will adopt this convention for our model without loss of generality. In practical applications, it is possible to use a different time lag or even a set of time lags, in which case the model could be extended to another additive form.

For a specific target region $s$, model~\eqref{eq:1} achieves two key objectives: 
\begin{inparaenum}
\item[(1)] determining the predictive relationship from other dimensions through the surfaces $\{\beta_{sg}(u,v)\}_{g=1}^S$, and 
\item[(2)] selecting a subset of significant predictors. 
\end{inparaenum}
These objectives contribute to the model's interpretability. However, a challenge arises when $S$ (the total number of regions) exceeds $n$ (the total number of time points), which is often the case in the HDFTS. To address this, it is essential to include only the significant coefficients in the additive model to ensure a more general and robust prediction model. Additionally, we aim to enhance the interpretability of each $\beta_{sg}(u,v)$ through local sparse estimation. For instance, in analyzing mortality rate curves for a region, we can identify region-specific and age-specific relationships between the mortality rates across different regions and the target region itself.

\subsection{Bivariate splines over triangulation}

One of our key contributions is retaining the functional form and using bivariate coefficient surfaces to represent the relationships between the predictors and the response. To effectively capture these relationships, it is essential to have a flexible representation of the surface coefficients. To this end, we employ bivariate smoothing splines, which also facilitate the two-fold variable selection problem -- both global and local selection -- ensuring the model remains interpretable. 

Bivariate splines are particularly well-suited for this purpose, as they allow us to partition the support of the surface into smaller, disjoint regions. In contrast, the conventional approach of representing a surface through the tensor product of univariate functions does not permit local variable selection, as modifying the surface in one region would affect the entire surface. Bivariate splines are more locally defined and enable us to set the surface to zero in specific regions without impacting the rest of the surface. Specifically, bivariate smoothing splines partition the support of the surface into triangles, a process referred to as triangulation, providing a more flexible and interpretable model. 

\begin{figure}[!htb]
\centering
\includegraphics[width=\textwidth]{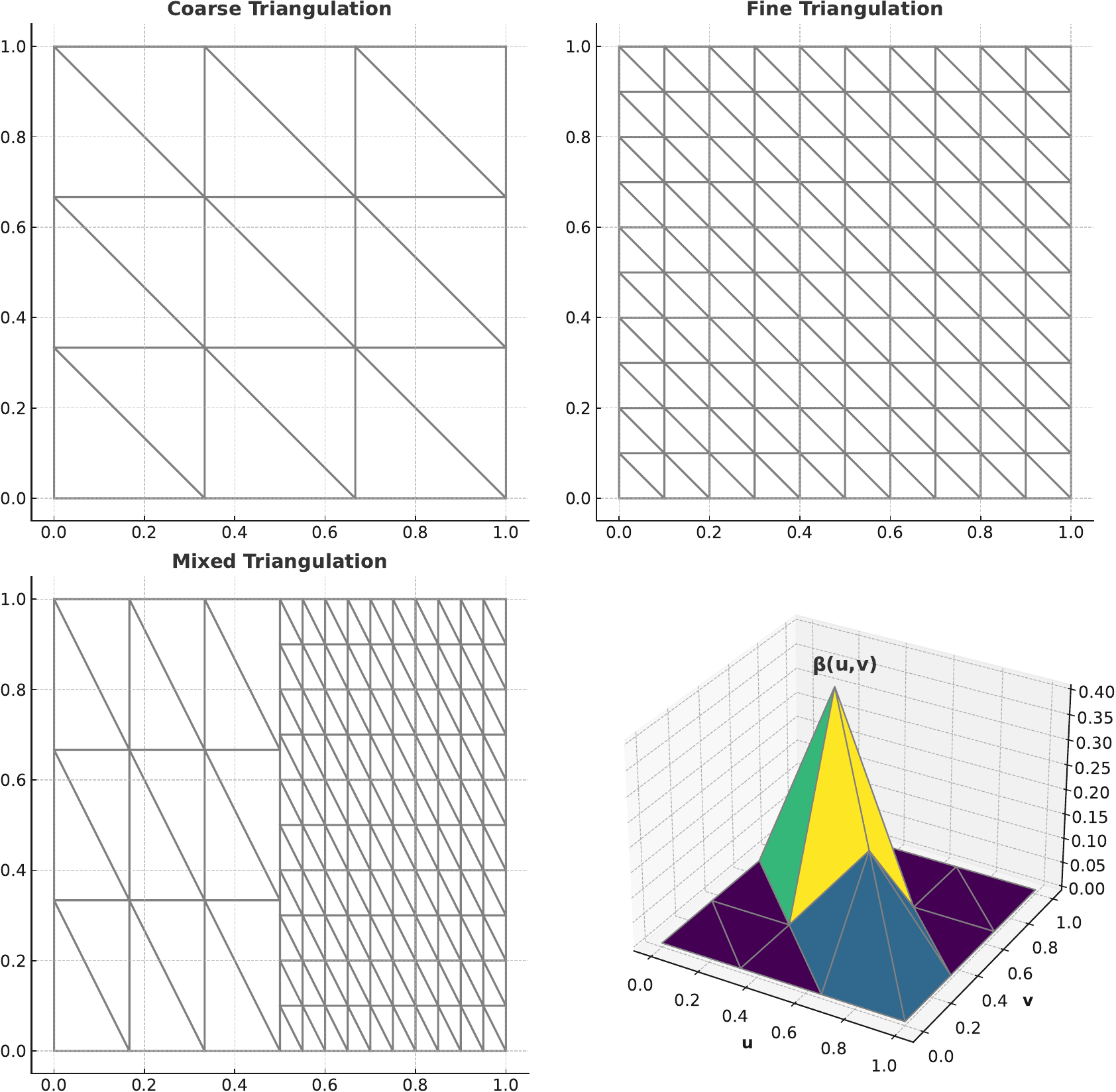}
\caption{Triangulation of the support $[0,1] \times [0,1]$. The top left plot demonstrates a coarse triangulation of 18 triangles. The top right plot shows a triangulation of the support with $200$ triangles that define a finer triangulation. The bottom left plot shows a mixture of both coarse and fine triangulation strategies. The bottom right plot shows how we can use triangulation to control where a surface could be zero and non-zero. The purple color indicates the surface is zero in a particular triangle, while other colors indicate the surface is non-zero in the corresponding triangle.}\label{fig:triangulation}
\end{figure}

As shown in Figure~\ref{fig:triangulation}, triangulation enables efficient partitioning of the surface's support. This approach provides greater control over the surface, particularly with finer triangulation, allowing for more precise adjustments and a more detailed representation.

We begin with the triangulation of the support $\mathcal{I} \times \mathcal{I}$ to define the surface coefficient. This allows us to represent the surface in terms of such a partition, namely, a set of triangles defined by their vertices. Let $l$ denote the index of individual triangles $A_{l}$ within the support $\mathcal{I} \times \mathcal{I}$. The triangulation $\Delta$ of $\mathcal{I} \times \mathcal{I}$ produces a collection of $L$ triangles, denoted $A_{1}, \dots, A_{l}, \dots, A_{L}$. For simplicity, we assume that the triangulation is the same for all coefficient surfaces $\beta_{sg}(u,v)$ for $g = 1, \dots, S$ associated with the $s$\textsuperscript{th} region.

Within each triangle, we define a set of bivariate basis functions $B_{ijk}(u,v)$, where $i + j + k = d$ for some integer degree $d \geq 1$. Unlike the tensor product method, the support of these basis functions $B_{ijk}(u,v)$ is defined over individual triangles rather than over the entire rectangular support. This approach provides more localized control over the surface, allowing for a more flexible and interpretable representation.

The fundamental construction of the coefficient surfaces uses the following basis expansion representation:
\begin{equation}
\beta_{sg}(u,v) = \sum_{l=1}^{L}\sum_{i+j+k=d}\gamma_{g,l,ijk}B_{ijk}(u,v),\label{eq:2}
\end{equation}
In~\eqref{eq:2}, the $\gamma_{g,l,ijk}$'s are the basis function coefficients. Furthermore, we let $\{B_{ijk}\}$'s be Bernstein polynomials defined over the triangle $A_{l}$. The exact form is 
\begin{equation*}
B_{ijk}(u,v) = \frac{d!}{i!j!k!}\left(\frac{u-a_1}{a_2-a_1}\right)^i \left(\frac{v-b_1}{b_2-b_1}\right)^j \left(1-\frac{u-a_1}{a_2-a_1}-\frac{v-b_1}{b_2-b_1}\right)^k,
\end{equation*}
where $i,j,k$ are non-negative integers such that $i+j+k = d$, and $(u,v)$ are coordinates within the triangle defined by arbitrary vertices $(a_1,b_1)$, $(a_2,b_2)$, and $(a_3,b_3)$ of the triangle $A_{l}$. The basis functions are defined over this triangle, and the coefficients $\gamma_{s^{'},l,ijk}$'s are the coefficients of the basis functions within the triangle. The number of basis functions in a triangle is equal to $\frac{(d+1)(d+2)}{2}$, which is also the number of coefficients $\gamma_{s^{'},l,ijk}$'s. This ensures that the model has a sufficient number of degrees of freedom to capture the relationships between the predictors and the response, while maintaining the flexibility needed for local variable selection.

Equation~\eqref{eq:2} forms a sieve estimator for the coefficient surface $\beta_{sg}(u,v)$. The degree $d$ is assumed to be finite and is a common assumption in the literature on sieve estimation \citep[e.g.,][]{Chen07}. By construction, a finite basis introduces an approximation error unless the true function $\beta_{sg}(u,v)$ lies exactly within the range of the chosen basis. In practice, we can choose a reasonable value for $d$ to balance the trade-off between the approximation error and the estimation variance. A value of $d = 3$ is a common choice. We have performed a sensitivity analysis on the choice of $d$ in our simulation studies (in Table~\ref{tab:degree_sensitivity_wide}), which shows that the results are not sensitive to the choice of $d$ as long as it is large enough.

\begin{figure}[!htb]
\centering
\includegraphics[width=\textwidth]{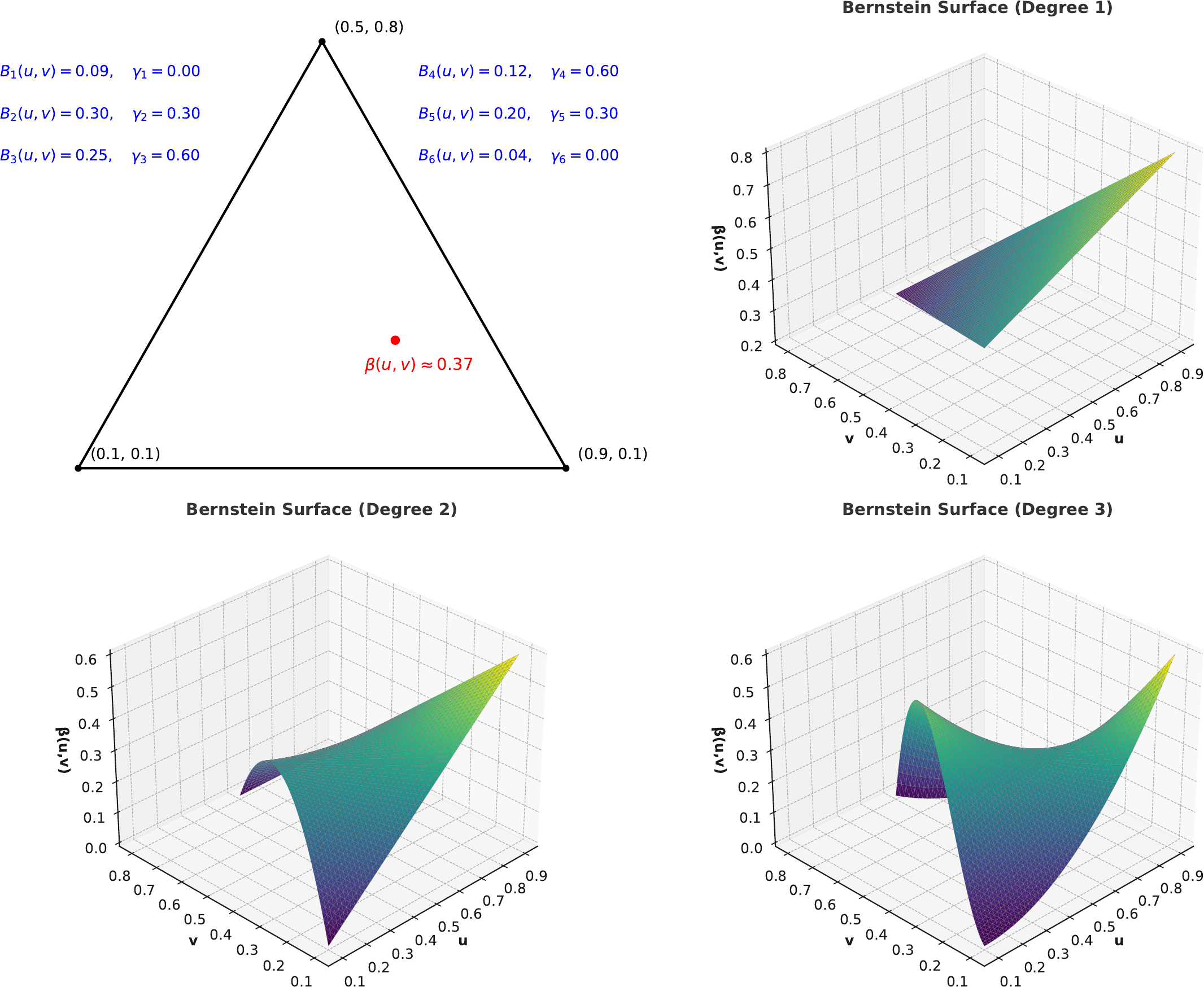}
\caption{The top left panel shows how to calculate the surface value for a point within the triangle defined by three vertices. The remaining plots demonstrate example surfaces expressed by different degrees for basis functions in this triangle.}\label{fig:berstein}
\end{figure}

For example, when $d = 2$, there are six basis functions within a triangle, each with its corresponding coefficient. Given a triangle $A_l$, the surface value at $(u,v) \in A_l$ is represented as a linear combination of the six basis functions, weighted by their respective coefficients. This is illustrated in the top-left plot of Figure~\ref{fig:berstein}. By increasing the degree of the basis functions, we can model more complex surfaces. Figure~\ref{fig:berstein} also presents some example surfaces within a triangle, each corresponding to different degrees of basis functions, highlighting the flexibility of the approach in capturing more intricate patterns.

For notational simplicity, we reindex the coefficients $\gamma_{g,l,ijk}$ as $\gamma_{g,l,q}$, where $q = 1, \dots, Q$, and $Q$ denotes the total number of basis functions within the triangle $A_l$. A key advantage of using the representation in~\eqref{eq:2}, as opposed to a tensor product of univariate basis functions, is its support for local sparse estimation. In essence, estimating the coefficient surfaces $\{\beta_{sg}(u,v)\}_{g=1}^S$ for each region~$s$ reduces to estimating the corresponding set of coefficients $\{\gamma_{g,l,q}\}$ for each predictor $g = 1, \dots, S$.

This formulation enables local sparsity: by setting all the coefficients associated with a particular triangle to zero, the surface becomes identically zero within that triangle. Moreover, if a surface $\beta_{sg}(u,v)$ is globally insignificant for a given predictor $g$, all its associated coefficients $\{\gamma_{g,l,q}\}$ can be set to zero across all triangles. This flexibility allows the model to adaptively eliminate both locally and globally irrelevant components, enhancing interpretability and reducing complexity. As shown in Figure~2, Bernstein polynomials over triangulations are specifically designed to approximate the two-dimensional surface with sparsity. This characteristic cannot be achieved by the B-spline or wavelet basis functions. That is, we partition the support of the coefficient into small and disjoint triangles where each triangle is associated with a set of basis functions that are non-zero only within that triangle. Hence, it is possible to achieve local sparsity by setting all coefficients associated with a triangle to zero. This is not possible with B-splines or wavelets, as their basis functions typically have overlapping supports that span multiple triangles. 

\subsection{Penalized smoothing bivariate splines}

The representation in~\eqref{eq:2} offers a flexible and convenient framework for constructing the coefficient surfaces. However, the estimation process must address several important objectives and constraints, namely, sparsity, continuity, and smoothness (or roughness control). These goals are critical to ensure both the interpretability and robustness of the model. To achieve these, we adopt a penalization approach, which allows us to systematically incorporate these aspects into the estimation procedure. The remainder of this section outlines the specific strategies and formulations used to enforce these desirable properties.

Sparsity constraint supports two key goals: selecting significant functional predictors (global sparsity); identifying regions where significant functional predictors influence the response (local sparsity). To achieve this, we apply a group bridge penalty on the coefficients interpolating the surface $\beta_{sg}$, promoting sparsity both across and within surfaces for improved interpretability and parsimony. The penalty is defined as follows:
\begin{equation}
\lambda_{1}(\sum_{l=1}^{L}c_{g,l}\|\bm{\gamma}_{g,l}\|^{\nu}_{1} + c_{g}||\bm{\gamma}_{g}||^{\nu}_{1}) ,\label{eq:grouplasso}
\end{equation}
where $\lambda_{1}$ is a sparsity parameter, $\bm{\gamma}_{g,l} = (\gamma_{g,l,1}, ..., \gamma_{g,l,Q})^{\top}$, $\bm{\gamma}_{g} = (\bm{\gamma}_{g,1}^{\top},..., \bm{\gamma}_{g,L}^{\top})^{\top}$, $^{\top}$ denotes matrix transpose, and $\nu \in (0,1)$. The weights $c_{g,l}$ and $c_{g}$ quantify the local and global contributions of each triangle and predictor, respectively, guiding the penalty structure to enhance sparsity and interpretability.

Equation~(3) contains two nested penalties that serve different purposes and are not redundant: a global predictor-level term, \(\lambda_{1}\,c_{g}\,\|\gamma_{g}\|_{1}^{\nu}\), which decides whether the entire surface for predictor \(g\) should be active at all; and local triangle-level terms,
\(\lambda_{1}\sum_{l=1}^{L} c_{g,l}\,\|\gamma_{g,l}\|_{1}^{\nu}\), which prunes or retains subregions within an active surface. We use separate weights \(c_{g}\) and \(c_{g,l}\) because the group sizes differ (the whole-surface block \(\gamma_{g}\) contains many more coefficients than a single triangle block \(\gamma_{g,l}\)). In practice, these weights are scaled by group cardinality (e.g., proportional to the square root of group size) to balance shrinkage across levels. In the reformulation~(7), the global criterion is included by treating \(\gamma_{g}\) as an additional \((L\!+\!1)\)-st group alongside the \(L\) triangle groups, so both levels are explicitly penalized. While group-bridge penalties with \(0<\nu<1\) can simultaneously encourage group selection and within-group sparsity at a single grouping level, our setting is nested (predictor-level surface versus triangle-level regions). Keeping both global and local terms prevents fragmentation (a predictor lingering due to a few tiny retained triangles) and yields clean predictor-level decisions together with sharper region-level sparsity within the surfaces.

The penalty term in~\eqref{eq:grouplasso} facilitates the selection of functional variables at both global and local levels, aligning with the group bridge approach introduced in \cite{groupbridge}. At the global level, the effect of each predictor, represented by $\beta_{s,g}$, is penalized through the $L_1$ norm of its associated coefficients $\bm{\gamma}_{g,l}$. When a predictor exhibits no substantial contribution, the corresponding surface can be shrunk entirely to zero, effectively removing that covariate from the model. This is achieved by grouping the coefficients according to the $g$\textsuperscript{th} predictor. At the local level, an additional layer of grouping, within each triangle, allows the model to enforce sparsity across regions of the surface. Specifically, the penalty term $\sum_{l=1}^{L}c_{g,l}\|\bm{\gamma}_{g,l}\|^{\nu}_{1}$ enables localized shrinkage, allowing surfaces to be zero in some regions while retaining nonzero values in others, thereby enhancing the model's interpretability and flexibility.


The second constraint ensures the continuity of coefficient surfaces across the triangulated domain. Although triangular basis functions enable localized modeling and are more flexible than tensor product bases, they require additional considerations to maintain smoothness across shared triangle edges. To address this, we impose linear constraints on the basis coefficients so that the interpolated surfaces from adjacent triangles join smoothly. Specifically, for each predictor $g$, we introduce a matrix $\bm{H}$, the rows of which are determined by the desired degree of smoothness and the columns of which correspond to the basis coefficients. The constraint $\bm{H} \bm{\gamma}_g = \bm{0}$ enforces continuity by requiring that the surfaces and their derivatives (to the desired degree) agree along shared edges. For further technical details on how to construct such constraints, see \cite{Lai_Schumaker_2007}.

The final constraint pertains to ensuring the smoothness of the coefficient surface. In the univariate spline smoothing context, a penalty function is commonly applied to prevent overfitting by controlling the roughness of the fitted curve. Similarly, we adopt this approach in the bivariate case, where the penalty is designed to regulate the smoothness of the estimated surface. The penalty function in this context is defined as follows:
\begin{equation*}
R_{\lambda_{2}}[\beta_{ss^{'}}(u,v)] = \lambda_{2} \int_{u,v\in \mathcal{I}}[D^2_{uu}\beta_{ss^{'}}(u,v)]^2+[D^2_{uv}\beta_{ss^{'}}(u,v)]^2 + [D^2_{vv}\beta_{ss^{'}}(u,v)]^2dudv,
\end{equation*}
where $D$ is the differential operator in the direction of $uu$, $uv$ and $vv$. By employing the penalization technique, we can simultaneously address these three constraints, thereby enhancing the interpretability and stability of the surface coefficients. In the subsequent section, we present the estimation algorithm for determining the surface coefficients.

\subsection{Estimation algorithm}\label{sec:2.3}

The primary objective function to estimate the surface coefficients $\{\bm{\gamma}_{g}\}_{g=1}^S$ for each target $s$\textsuperscript{th} region is based on the least squares criterion. Specifically, we aim to estimate the surface coefficients by minimizing the squared distance between the functional response and the additive form of the predictors, as expressed in the following equation:
\begin{equation*}
\argmin_{\{\bm{\gamma}_{g}\}_{g=1}^{S}}\sum_{t=2}^{n}\left\|\X_{ts}(\cdot) - \sum_{g=1}^S\int_{u\in\mathcal{I}}\beta_{sg}(u,\cdot)\X_{t-1,g}(u)du\right\|^2,
\end{equation*}
where $\|\cdot\|$ is the functional norm in $L_2(\mathcal{I})$, $\bm{\gamma}_{g}=(\gamma_{g,l = 1, q = 1},\dots, \gamma_{g, l = L, q = Q})^{\top}$, which is a $L \times Q$ \textcolor{blue}{vector}. Let $n$ denote the index of the last functional observation across the panel. Incorporating the penalization terms, the updated objective function for estimating the surface coefficients is then expressed as follows:
\begin{align}
\mathcal{L}_{n}(\bm{\gamma}) = \sum_{t=1+\delta}^{n}\left\|\X_{ts}(\cdot) - \sum_{g=1}^S\int_{u\in\mathcal{I}}\beta_{sg}(u,\cdot)\X_{t-1,g}(u)du\right\|^2&+\lambda_{1}\sum_{g=1}^{S}(\sum_{l=1}^{L}c_{g,l}\|\bm{\gamma}_{g,l}\|^{\nu}_{1} + c_{g}||\bm{\gamma}_{g}||^{\nu}_{1})\notag\\
&+ \sum_{g = 1}^{S} R_{\lambda_{2}}[\beta_{sg}(u,v)]\label{eq:4}
\end{align}
subject to $\bm{H}\bm{\gamma}_{g}=0, \forall g$.

The objective function in~\eqref{eq:4} has three components:
\begin{inparaenum}
\item[(1)] the squared distance objective;
\item[(2)] the sparsity penalty; and
\item[(3)] the roughness penalty
\end{inparaenum} 
with a linear constraint. For simplicity, we assume that each individual curve has sufficient observation points $\{u_{m}\}_{m=1}^{M}$ such that the integral $\int_{u \in \mathcal{I}}$ can be accurately approximated using the Riemannian summation. Alternatively, quadrature methods, such as the trapezoidal or Simpson's rule, can also be employed for approximating integrals. In addition, it is assumed that the discrete evaluation points, $u_m$'s, are consistent across all individual functional data for simplicity. Given the target functional time series $\X_{ts}(u)$ for $t = 1, \dots, n$ and the predictors $\X_{t - \delta, g}(u)$ for $g = 1, \dots, S$, the first component of the objective function can be expressed in quadratic form: 
\begin{align}
\sum^n_{t=1+\delta}\Big\|\X_{ts}(\cdot) - \sum_{g=1}^S\int_{u\in\mathcal{I}}\beta_{sg}(u,\cdot)\X_{t-\delta,g}(u)du\Big\|^2  &=\sum^n_{t=1+\delta}\int \Big[\X_{ts}(v) - \sum^{S}_{g=1}\int_{u\in\mathcal{I}}\beta_{sg}(u,v)\X_{t-\delta,g}(u)du\Big]^2 dv \quad \notag\\
&\approx \frac{1}{M}\sum^n_{t=1+\delta}\sum^M_{m=1}\Big[\X_{ts}(v_m) - \sum^{S}_{g=1}\int_{u\in\mathcal{I}}\beta_{sg}(u,v_m)\X_{t-\delta,g}(u)du\Big]^2\notag\\
&= (\bm{y} - \bm{\Psi}\bm{\gamma})^{\top}(\bm{y} - \bm{\Psi}\bm{\gamma}) \label{eq:5}
\end{align}
where $\bm{y} = [\X_{1+\delta,s}(v_1),\dots,\X_{T,s}(v_1), \dots,\dots,\X_{1+\delta,s}(v_M),\dots,\X_{T,s}(v_M)]^{\top}$. In quadratic form~\eqref{eq:5}, $\bm{\Psi} = (\bm{\Psi}_{1}, \dots, \bm{\Psi}_{S})$, where $\bm{\Psi}_{g}$ is a $M \times (L \times Q)$ matrix. Specifically, each row of $\bm{\Psi}_{g}$ consists of the integrals $\int_{u\in\mathcal{I}} B_{l,q}(u, v_m) \X_{t-\delta,g}(u) \, du$ for $l = 1, \dots, L$ and $q = 1, \dots, Q$, with $m$ ranging from~$1$ to $M$. The coefficient vector is defined as $\bm{\gamma} = (\bm{\gamma}_{1}^{\top}, \dots, \bm{\gamma}_{S}^{\top})^{\top}$, where $\bm{\gamma}_{g}$ represents the set of coefficients for the $g$\textsuperscript{th} predictor.

The roughness penalty $\sum_{g = 1}^{S} R_{\lambda_{2}}[\beta_{sg}(u,v)]$ can be expressed in quadratic form as $\lambda_{2} \bm{\gamma} \bm{R} \bm{\gamma}$, where $\bm{R}$ is a block diagonal matrix of size $(S \times (L \times Q)) \times (S \times (L \times Q))$, with each block $\bm{R}_{g}$ being a $(L \times Q) \times (L \times Q)$ matrix corresponding to the $g$\textsuperscript{th} predictor. For simplicity and without loss of generality, we assume that the triangulation of all coefficient surfaces is the same, resulting in identical matrices $\{\bm{R}_{g}\}_{g = 1}^{S}$.

The sparsity penalty $\lambda_{1} (\sum_{l=1}^{L} c_{g,l} \|\bm{\gamma}_{g,l}\|_1^{\nu} + c_{g} \|\bm{\gamma}_{g}\|_1^{\nu})$ resembles a grouped variable selection problem, where the coefficients are grouped by their respective triangles and predictor $g$. Due to the presence of the sparsity penalty and the power term $0 < \nu < 1$, the objective function~\eqref{eq:4} is non-convex. To address this issue, we adopt the approach outlined in \cite{groupbridge} and reformulate the objective function equivalently as follows:
\begin{equation}
(\bm{y} - \bm{\Psi}\bm{\gamma})^{\top}(\bm{y} - \bm{\Psi}\bm{\gamma})+\lambda_2 \bm{\gamma} \bm{R}\bm{\gamma} + \sum^S_{s^{'}=1}\sum_{l=1}^{L+1}\theta_{s^{'},l}^{1-1/\nu}c_{s^{'},l}^{1/\nu}\|\bm{\gamma}_{g,l}\|_1+\tau\sum^S_{s^{'}=1}\sum_{l=1}^{L+1}\theta_{s^{'},l},\label{eq:6}
\end{equation}
with a set of newly introduced parameters $\bm{\theta} = \{\theta_{s^{'},l}\}$'s with each $\theta$ being non-negative and $\tau = [\lambda_1\nu^{\nu}(1-\nu)^{1-\nu}]^{1/(1-\nu)}$. For consistency and simplicity in notation, we represent $||\bm{\gamma}_{g}||_{1}$ as $||\bm{\gamma}_{g,L+1}||_{1}$. In doing so, the summation from $1$ to $L+1$ will encompass the global penalty. The term $L+1$ represents the global penalty, while the first terms $L$ represent the local penalties for each triangle within the triangulation. To obtain the minimizer $\widehat{\bm{\gamma}}$ of~\eqref{eq:4}, it is sufficient to minimize the objective function~\eqref{eq:6} with respect to the parameters $(\widehat{\theta}, \widehat{\bm{\gamma}})$. 

To address the roughness penalty and smoothness constraints on the bivariate basis functions, we use the concatenation of the design matrix. Specifically, we define a new design matrix $\bm{\Psi}^{*}$ as the vertical concatenation of $\bm{\Psi}^{\top}$, $\bm{H}^{\top}$, and $\omega \bm{R}^{1/2}$, such that $\bm{\Psi}^* = (\bm{\Psi}^{\top}, \bm{H}^{\top}, \omega \bm{R}^{1/2})^{\top}$, where $\omega=\sqrt{\lambda_2}$. Additionally, we extend the vector $\bm{y}$ to $\bm{y}^*$ to ensure that it matches the length of the new design matrix $\bm{\Psi}^*$.

For the selection of the group variable on $\bm{\gamma}$, the penalized estimate is achieved by transforming the original coefficients. For each $g$, define an $(L+1) \times (L+1)$ matrix with diagonal elements $w_{g,l}^{-1}$, where $w_{g,l}^{-1} = \sum_{l=1}^{L+1} \theta_{g,l}^{1 - 1/\nu}c_{g,l}^{1/\nu}$, representing the sum of the weights from the $l$\textsuperscript{th} triangle and the entire region $g$. Using this transformation, we define $\gamma^{*}_{g} = \bm{W}^{-1} \gamma_{g}$, where $\bm{W}$ denotes the matrix of block elements $w_{g,l}$'s with $g$ being the block and $l$ being the diagonal element in that block. This allows us to rewrite the objective function~\eqref{eq:6} in terms of these new quantities
\begin{equation}
(\bm{y}^*-\bm{\Psi}^*\bm{\gamma}^*)^{\top}(\bm{y}^*-\bm{\Psi}^*\bm{\gamma}^*)+ \sum^S_{g=1}\sum_{l=1}^{L+1}\|\bm{\gamma}^{*}_{g,l}\|_1+\tau\sum^S_{g=1}\sum_{l=1}^{L+1}\theta_{g,l}. \label{eq:8}
\end{equation}

We can employ a heuristic algorithm to estimate the additive model. In each iteration, we consider a reduced version of the objective function~\eqref{eq:8} by fixing $g$ to a single predictor. This simplification reduces the number of columns in the design matrix $\bm{\Psi}^{*}$ and the number of coefficients in $\bm{\gamma}^{*}$ and $\bm{\theta}^{*}$. The fitting algorithm proceeds as follows, with the iteration index denoted as $^{(\cdot)}$:
\begin{itemize}
\item[Step 1] Initialization: Obtain initial estimates $\bm{\gamma}_{g}^{(0)}$ without the sparsity penalty for $g = 1,..., S$.
\item[Step 2] At each iteration, shuffle the order of $g = 1,..., S$ and loop over shuffled $g$ from $1$ to $S$:
\begin{itemize}
\item[Step 2a] Start with the original observation vector $\bm{y}$. Extend $\bm{y}^* = (\bm{y}^{\top},\bm{0}^{\top})^{\top}$ to match the number rows of $\bm{\Psi}^*$.
\item[Step 2b] Minimize $(\bm{y}^*-\bm{\Psi}^*\bm{\gamma}_{g}^*)^{\top}(\bm{y}^*-\bm{\Psi}^*\bm{\gamma}_{g}^*)+ \sum_{l}\|\bm{\gamma}^{*}_{g,l}\|_1+\tau\sum_{l}\theta_{s^{'},l}$ with respect to $(\bm{\gamma}_{g}, \bm{\theta}_{g})$.
\item[Step 2c] Calculate the residual $(\bm{y} - \bm{\Psi}\bm{\gamma}^{(\text{iter})}_{g})$ and update the new observation vector $\bm{y}$ with the calculated residuals.
\item[Step 2d] Repeat step a-c for $g = 1,..., S$.
\end{itemize}
\item[Step 3] Repeat Step~2 until convergence.
\item[Step 4] After finding the significant subset of $\{\beta_{sg}\}$'s, we refit the model with only the significant ones and without the sparsity penalty.
\end{itemize}
\noindent During the minimization process, any coefficients that have been shrunk to zero are removed from the iterative procedure.

Given that we have employed an additive model in Equation~\eqref{eq:1}, identifiability issues may arise. To address this, we can standardize the functional data, which helps mitigate such concerns. Additionally, we can modify the uniform triangulation $A_l$ into a set of region-specific triangulations $\{A_{g,l}\}_{l=1}^{L}$ for each distinct $g$\textsuperscript{th} predictor. This ensures that the coefficient surfaces $\beta_{sg}(u,v)$ for different predictors do not overlap, thereby resolving potential identifiability issues.

The weights $c_{g,l}$ can be chosen in proportion to the number of coefficients within the associated group, for example, setting $c_{g,l} \propto Q$. In addition, several hyperparameters must be selected: $d$, $L$, $\lambda_{1}$, and $\lambda_{2}$. Sensitivity analysis (in Table~\ref{tab:degree_sensitivity_wide}) conducted on the choice of $d$ in our simulation studies, shows that the results are not sensitive to the choice of $d$ as long as it is large enough. In fact, we found that $d = 3$ is sufficient for our purposes and there is no significant difference beyond $d = 5$. However, the number of triangles, $L$, can be chosen based on the data size and the desired complexity of the surface. A higher value of $L$ allows the fitting of a more complex surface but increases the computational cost.

The primary focus is on selecting the regularization parameters $\lambda_{1}$ and $\lambda_{2}$. To tune these parameters, we employ a training-validation split, which is particularly suitable given the longitudinal nature of the HDFTS. We define a grid of candidate values for $\lambda_{1}$ and $\lambda_{2}$, such as $10^{-5}, 10^{-4}, 10^{-3}, 10^{-2}, 10^{-1}$ for both. The tuning grid is the combination of these two sets. For tuning and model evaluation, we consider splitting the data by 60\%-20\%-20\% into training-validation-testing sets, respectively. The prediction performances for observations in the validation set are used to select the optimal values for $\lambda_{1}$ and $\lambda_{2}$, which minimize the mean squared prediction errors. The final model is then fitted with both training and validation data and the chosen values of $\lambda_{1}$ and $\lambda_{2}$. In the end, we will examine the actual forecast performance of the model in the test set and compare it with other benchmark models.

\section{Theoretical results}\label{sec:theo}

We let $\X_{g}$ represent the functional variable in each predictor region $g = 1,..., S$ and observed at a sequence of time points $t = 1,..., n$ with $X_{tg}$. Without loss of generality and for a target region $s = 1,..., S$, we assume that the set of functional variables $\{\X_{g}\}_{g=1}^{S}$ is ordered such that $\beta_{sg} \neq 0$ for $g = 1, \dots, J_{1}$ and $\beta_{sg} = 0$ for $g = J_{1}+1, \dots, S$. A functional coefficient $\beta_{sg}$ is considered zero if and only if $\beta_{sg}(u,v) = 0$ for all $(u, v)$. We define the sets $B_1 : \{g = 1, \dots, J_1\}$ and $B_2 : \{g = J_1+1, \dots, S\}$ to represent the groups of functional predictors that are globally active and inactive, respectively. This notion can be directly applied to the coefficient vectors $\bm{\gamma}_g$, such that, for example, $\bm{\gamma}_{g \in B_1} \neq \bm{0}$. 

To derive the theoretical properties of the estimator $\widehat{\bm{\gamma}}$ obtained from minimizing the objective function~\eqref{eq:4}, that is,
\begin{equation*}
\widehat{\bm{\gamma}}= \underset{\bm{\gamma}}{\argmin} ||\bm{y} - \bm{\Psi}\bm{\gamma}||^{2}_{2} + \lambda_{1}\sum_{g=1}^{S}\sum_{l=1}^{L+1}c_{g,l}||\bm{\gamma}_{g,l}||^{\nu}_{1} + \lambda_{2}\bm{\gamma}^{\top}\bm{R}\bm{\gamma}
\end{equation*}
hence the estimator of $\beta_{sg}$'s has the form of $\widehat{\beta}_{sg}(u,v) = \sum_{l=1}^{L}\sum_{q=1}^{Q} \widehat{\gamma}_{g,l,q} B_{l,q}(u,v)$, we make the following assumptions:
\begin{assumption}\label{assumption:1}
The true coefficient surfaces $\beta_{sg}$'s are smooth with order $m$ and bounded, i.e., $\sup_{u,v \in \mathcal{I}}|\bm{\beta}_{g}(u,v)| <~\infty$.
\end{assumption}

\begin{assumption}\label{assumption:2}
The Ridge matrix $\bm{\Psi}^{\top}\bm{\Psi} + \lambda_{2}\bm{R}$ has the the minimum and maximum eigenvalues satisfy $a \leq \rho_{\min} < \rho_{\max} \leq b$ for some constants $0 <a < b < \infty$.
\end{assumption}

\begin{assumption}\label{assumption:3}
The penalty parameters $\lim_{n \rightarrow \infty}\lambda_{1}/n = 0$ and $\lim_{n \rightarrow \infty}\lambda_{2}/n = 0$.
\end{assumption}

\begin{assumption}\label{assumption:4}  
Let $\bm{R}$ be the penalty matrix for the penalty term $\mathcal{P}_{2}(\cdot; \lambda_{2})$ in~\eqref{eq:6}. The eigenvalues of $\bm{R}$ are bounded away from zero, i.e., $\lambda_{\min}(\bm{R}) \geq c > 0$.
\end{assumption}

\begin{assumption}\label{assumption:5}
Let $c_{\max}  = \max(\{c_{g}\}_{g=1}^{S})$, the ratio $\frac{\rho_{\max} + \lambda_{2}}{c_{\max}(1 - \nu)\lambda_{1}} \rightarrow 0$ as $nK \rightarrow \infty$.
\end{assumption}

\begin{assumption}\label{assumption:6}
The error functions $\epsilon_{ts}(v)$ are independent and identically distributed with mean zero and variance $\sigma^{2} < \infty$.
\end{assumption}

\begin{assumption}\label{assumption:7}
Let $\Omega\subset\mathbb{R}^2$ be the domain and let
$\{A_L\}_{L\ge1}$ be a sequence of triangular meshes of $\Omega$.
Denote by $h_l:=\operatorname{diam}(l)$ the diameter of a triangle $l \in A_L$, $h_l:=\max_{l\in A_L} h_l$, and $\rho(l)$ the in-radius of~$l$.
Assume the meshes are
(i) \emph{shape-regular}: there exists $\kappa>0$ such that $\rho(l)\ge \kappa\, h_l$
for all $l\in A_L$ and all $l$, and
(ii) \emph{quasi-uniform}: there exists $\theta\ge1$ such that
$\max_{T\in\Delta_L} h_T / \min_{l\in A_L} h_l \le \theta$ for all $l$.
\end{assumption}

Given HDFTS $\{X_{ts}(u)\}_{s=1}^{S}$ and the model~\eqref{eq:1}, we can first show that the estimator $\widehat{\bm{\gamma}}$
is consistent and accompanied by the convergence rate as follows: 
\begin{theorem}\label{thm:1} 
Let $\bm{\gamma}_{\star}$ be the true coefficients in generating the discrete observations $\bm{y} = \bm{\Psi}\bm{\gamma}_{\star} + \bm{\epsilon}$, where $\bm{\epsilon}$ is the discrete evaluation of the error term in Equation~\ref{eq:1}. Let $\widehat{\bm{\gamma}}$ be the estimated coefficients from the optimization of the objective function~\eqref{eq:4}. There exist some constants $0 < a < b < \infty$, $\eta_{\gamma} < \infty$, and $K = S\times L \times Q$. Then, we have
\begin{equation*}
\mathbb{E}(||\widehat{\bm{\gamma}} - \bm{\gamma}_{\star}||^{2}_{2}) \leq  4\frac{\lambda^{2}_{1}\eta^{2}_{\gamma} + \lambda_{2}^{2}||\gammatrue||_{2}^{2} + nMb\sigma^{2}}{(nMa + \lambda_{2})^{2}},
\end{equation*}
where $\sigma^2<\infty$ is the variance of the error term $\bm{\epsilon}$ with Assumptions 1 - 6.
\end{theorem}

It is possible that the representation in~(\ref{eq:2}) may not perfectly represent the true $\beta_{sg}(u,v)$, leading to an approximation error. If this is the case, we consider the true surface to be $\beta^{\circ}_{sg}$ which can be written as 
\begin{equation*}
\beta^{\circ}_{sg}(u,v) = \beta^{\star}_{sg} + \mathfrak b_{sg}(u,v) \quad\text{or}\quad\sum_{l=1}^{L} \gamma^{\star}_{sg,l} b_l(u,v) + \mathfrak b_{sg}(u,v),
\end{equation*}
where $\mathfrak b_{sg}(u,v)$ is the deterministic approximation error term.

Theorem~1 shows the convergence of an estimate $\widehat{\beta}_{sg}$ to the best approximation of the true (best-in-class) surface $\beta^{\star}_{sg}$ within the chosen basis. The following lemma discusses the convergence of $\widehat{\beta}_{sg}$ to $\beta^{\circ}_{sg}$, considering the approximation error incurred when the basis cannot capture the true surface. Here, we consider $\beta^{\circ}_{sg}$ as the ground-truth surface for~(\ref{eq:1}) and supposedly an infinite-dimensional object.

\textbf{Lemma~1.} \textit{Given Theorem 1 holds and additional Assumption 7, we have the consistency of coefficients $\widehat{\bm{\gamma}}$ to $\bm{\gamma}_{\star}$ as $n\to\infty$. For a target region $s$ and a predictor $g$, we assume that we have $||\widehat{\bm{\gamma}}_{g} -  \bm{\gamma}_{\star, g}|| = O_p(\eta_n)$, where $\eta_n\to 0$ as $n\to\infty$. Then, we have}
\begin{equation*}
\left\|\widehat{\beta}_{sg} - \beta^{\star}_{sg}\right\|_{\mathcal{L}_2} = O_p(\eta_n) + ||\mathfrak b_{sg}||_{\mathcal{L}_2}.
\end{equation*}
with
\begin{equation*}
\|\mathfrak b_{sg}\|_{\mathcal{L}_2} \leq C h_{L}^{\min\{d+1,m\}},
\end{equation*}
\textit{where $C$ is a positive constant, $h_L$ is the maximum diameter of the triangles in the triangulation, $d$ is the degree of the polynomial basis, and $m$ is the smoothness of the true surface $\beta^{\circ}_{sg}$. Consequently, if $h_L\to 0$ as $n\to\infty$, then we have the consistency of $\widehat{\beta}_{sg}$ to $\beta^{\star}_{sg}$ in $\mathcal{L}_2$ norm.} This lemma is similar to the discussion on the approximation power of Bernstein polynomials in \cite{Lai_Schumaker_2007}.

Under the consistency of estimation, as in Theorem~\ref{thm:1}, we can establish the oracle property of our estimator at two levels: global and local. The global level ensures the consistency of selecting significant predictors in $B_{1}$ and then we can derive the asymptotic normality of the estimated coefficients $\bm{\gamma}_{g}$ for $g \in B_1$. That is, 
\begin{theorem}\label{thm:2}  Global oracle property. With Assumptions 1-6, we can show that the probability of correctly identifying the significant predictors in $B_{1}$ and excluding the insignificant ones in $B_{2}$ approaches 1 as $n \rightarrow \infty$. That is,
\begin{equation*}
P(\widehat{\bm{\gamma}}_{g \in B_{2}} = \bm{0}) \rightarrow 1 \quad \text{or} \quad P(\widehat{\beta}_{sg} = 0, \forall g \in B_{2}) \rightarrow 1.
\end{equation*}
Furthermore, for the predictors in the active set $B_{1}$, their estimated coefficients $\widehat{\bm{\gamma}}_{g}$ converge to the true coefficients $\gammatrue$ at the rate of $\sqrt{nM}$ and follow an asymptotic normal distribution:
\begin{equation*}
  \sqrt{nM}(\widehat{\bm{\gamma}}_{B_{1}} - \gammatrue) \xrightarrow{d} \mathcal{N}(0, \sigma^{2}\bm{\Sigma}_{B_{1}}^{-1}).
\end{equation*}
\end{theorem}
\noindent where $\bm{\Sigma}_{B_{1}} = (\bm{\Psi}_{B_{1}}^{\top}\bm{\Psi}_{B_{1}} + \lambda_{2}\bm{R}_{B_{1}})^{-1}\bm{\Psi}_{B_{1}}^{\top}\bm{\Psi}_{B_{1}}(\bm{\Psi}_{B_{1}}^{\top}\bm{\Psi}_{B_{1}} + \lambda_{2}\bm{R}_{B_{1}})^{-1}$. This selection process can be effectively facilitated by the penalty term with $l = L+1$ in the objective function in~\eqref{eq:6}. This penalty on all coefficients from a single predictor helps us with global selection of functional covariates. 

The global oracle property ensures that our method can correctly identify all significant dimensions in the panel with high probability. That is, if a dimension $g$ is significant in predicting the response at the location $s$, our method will select it with high probability as the sample size increases. Equivalently, if a dimension $g$ is not significant to predict the response at the location $s$, our method will exclude it with high probability as the sample size increases. Given that a dimension $g$ to $s$ is significant, Theorem~2 also guarantees that the estimated coefficient surface $\widehat{\beta}_{sg}$ converges to the best-in-class coefficient surface $\beta^{\star}_{sg}$ at the optimal rate.

Based on the global oracle property, that is, selecting significant $g$'s to predict a target region $s$, we can further investigate the local oracle property of the estimator $\widehat{\bm{\gamma}}_{g}$ for each selected $g \in B_{1}$. This helps us to identify not only significant predictors, but also significant contributions within those predictors. To establish a theoretical guarantee for the local oracle property of the estimator $\widehat{\bm{\gamma}}_{g}$ for every $g \in B_{1}$, we consider the triangulation, a set of triangles $\{A_{l}\}_{l=1}^{L}$, which can be partitioned (and reordered if necessary) into two disjoint sets, $C^{g}_{1}: \{l = 1, \dots, J^{g}_{2}\}$ and $C^{g}_{2}: \{J^{g}_{2}+1, \dots, L\}$. For each $l \in C^{g}_{2}$, we assume that $\beta_{sg}(u,v) = 0 \hspace*{0.5em}\forall (u,v) \in A_{l}$. This can be equivalently viewed as partitioning the coefficient vector $\bm{\gamma}_{g}$ into $(\bm{\gamma}^{\top}_{C^{g}_{1}}, \bm{0})^{\top}$, where $\bm{\gamma}_{C^{g}_{1}} = (\bm{\gamma}^{\top}_{g,1}, \dots, \bm{\gamma}^{\top}_{g,J^{g}_{2}})^{\top}$ consists of the coefficients corresponding to the basis functions of the triangles in $C^{g}_{1}$. 

Consequently, the local oracle property of the estimator $\widehat{\beta}_{sg}(u,v)$ is then defined through the behavior of $\bm{\gamma}_{g}$ as follows:
\begin{theorem} \label{thm:3} Local oracle property. With Assumptions 1-6 and Theorem 2, we can show that for each selected predictor $g \in B_{1}$, the probability of correctly identifying the significant triangles in $C^{g}_{1}$ and excluding the insignificant ones in $C^{g}_{2}$ approaches 1 as $n \rightarrow \infty$. That is,
\begin{equation*}
P(\widehat{\bm{\gamma}}_{g,l \in C^{g}_{2}} = 0) \rightarrow 1,  \forall g \in B_{1} \quad \text{or} \quad P(\widehat{\beta}_{sg}(u,v) = 0, \forall (u,v) \in A_{l}, l \in C^{g}_{2}) \rightarrow 1,  \forall g \in B_{1}.
\end{equation*}
Furthermore, for the triangles in the active set $C^{g}_{1}$, their estimated coefficients $\widehat{\bm{\gamma}}_{g,l \in C^{g}_{1}}$ converge to the true coefficients $\gammatrue$ at the rate of $\sqrt{nM}$ and follow an asymptotic normal distribution:
\begin{equation*}\sqrt{nM}(\widehat{\bm{\gamma}}_{g, l \in C^{g}_{1}} - \gammatrue) \xrightarrow{d} \mathcal{N}(0, \sigma^{2}\bm{\Sigma}_{g,C_{1}}^{-1}).
\end{equation*}
where $\bm{\Sigma}_{g,C_{1}} = (\bm{\Psi}_{g,C_{1}}^{\top}\bm{\Psi}_{g,C_{1}} + \lambda_{2}\bm{R}_{g,C_{1}})^{-1}\bm{\Psi}_{g,C_{1}}^{\top}\bm{\Psi}_{g,C_{1}}(\bm{\Psi}_{g,C_{1}}^{\top}\bm{\Psi}_{g,C_{1}} + \lambda_{2}\bm{R}_{g,C_{1}})^{-1}$. 
\end{theorem}  

Beyond the global oracle property, Theorem~3 establishes the local oracle property. This property ensures that our method can correctly identify all significant triangles within each selected dimension $g \in B_{1}$ with high probability. That is, if a triangle $l$ within a selected dimension $g$ is significant for predicting the response at location $s$, our method will select it with high probability as the sample size increases. Equivalently, if a triangle $l$ within a selected dimension $g$ is not significant for predicting the response at location $s$, our method will exclude it with high probability as the sample size increases. Since the support of the surface is partitioned into triangles, the local oracle property ensures that our method can correctly identify where the coefficient surface $\beta_{sg}(u,v)$ is significant with probability tending to 1 as the sample size increases.

The proof of Theorems~\ref{thm:1},~\ref{thm:2}, and~\ref{thm:3} are provided in the~Appendix with necessary assumptions.

\section{Monte Carlo simulation studies}\label{sec:simu}

Simulation studies are conducted to evaluate the performance of the proposed model in estimating the coefficient surfaces and predicting the HDFTS. Forecast performance is assessed using two key metrics: the finite sample mean absolute forecast error (MAFE) and the mean square forecast error (MSFE) for HDFTS. Specifically, the focus of the simulation studies is to investigate how the two levels of sparsity (global and local) contribute to improved prediction and estimation, in comparison to scenarios where either no sparsity or only global sparsity is applied to the coefficient surfaces.

\subsection{Data and model-generating processes}

The simulation studies begin by generating the HDFTS $X_{ts}(v)$ for $t = 1, \dots, n$, $s = 1, \dots, S$, where $S = 7$ and $n \in \{50, 100, 200, 500\}$ and each individual curve has sufficient observation points $\{v_{1},\dots, v_{M=50}\}$. For each time series length $n$, we repeat the Monte Carlo simulation 1,000 times. 
For each dimension $s = 1, ..., S$ in the HDFTS, we first generate a self-regressive functional time series $\{\X_{ts}\}$ from a FAR(1) model with a random coefficient surface $\beta_{ss}(u,v) = \Gamma_{s}(u,v)$ as in the following model. This helps us to ensure that the spectral norm of the corresponding kernel operator is less than~1.
\begin{equation*}
\X_{ts}(v) =\int_{u\in \mathcal{I}}\Gamma_{s}(u,v)\X_{t-1,s}(u)du + \omega_{ts}(v),
\end{equation*}
where 
\begin{equation*}
\Gamma_{s}(u,v) = C_{s}\text{exp}(-\frac{(u+v)^{2}}{2})
\end{equation*}
for some constant $C_{s} \in [0,1]$ and $\omega_{ts}(v)$ is a zero-mean error function with variance of $1$. We then proceed by simulating the target functional time series $\X_{ts}(v)$ using the following additive model:
\begin{equation}
\X_{ts}(v) = \sum_{g=1}^{S}\int_{u \in \mathcal{I}}\beta_{sg}(u,v)X_{t-1,g}(u)du + \epsilon_{ts}(v),\quad S = 7,\label{eq:9}
\end{equation}
which would ensure each dimension is linked to others and the condition for stationarity is satisfied. The coefficients surfaces $\beta_{sg}(u,v)$ for each $s$ are chosen as follows. First, we set $\beta_{ss}$ as a fully specified coefficient without sparse regions, identical to $\Gamma_{s}(u,v)$. The remaining coefficients will be partially sparse. These coefficient surfaces are illustrated in Figure~4. Finally, we incorporate a group of completely sparse coefficient surfaces into the fitting process, thereby increasing the model complexity and involving a total of seven predictors for $g = 1,..., 7$.
\begin{figure}[!htb]
\centering
\includegraphics[width=\textwidth]{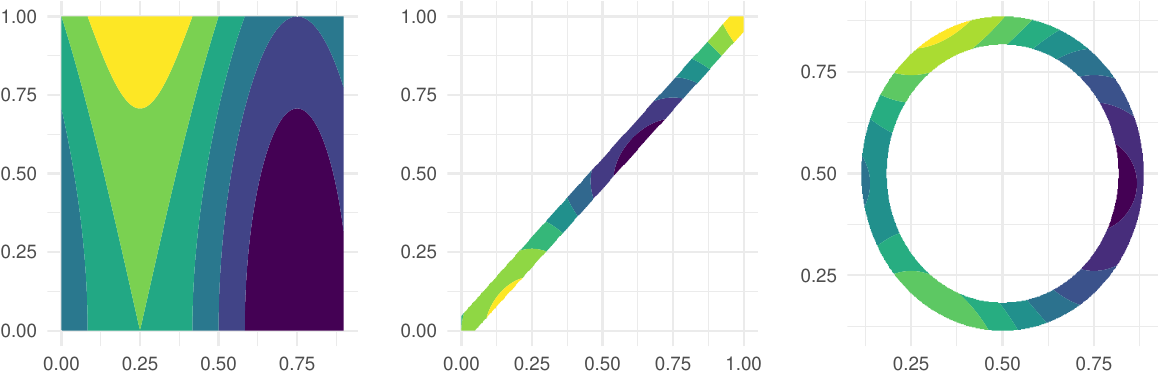}
\caption{Partially sparse coefficient surfaces, with the white region indicating the surface is zero. Each column corresponds to a differently shaped coefficient surface.}\label{fig:coefsurfaces}
\end{figure}

For each Monte Carlo simulation of the HDFTS, we reserve the last 20\% of the series as the testing data to compute the Mean Absolute Forecast Error (MAFE) and Mean Squared Forecast Error (MSFE), defined as follows:
\begin{align}
\text{MAFE} &= \frac{1}{n^{\prime}}\frac{1}{S} \sum_{t^{\prime}=1}^{n^{\prime}}\sum_{s=1}^{S}\int_{v\in \mathcal{I}}\big|\X_{t^{\prime}s}(v) - \widehat{\X}_{t^{\prime}s}(v)\big|dv,\label{eq:MAFE}\\
\text{MSFE} &= \frac{1}{n^{\prime}}\frac{1}{S} \sum_{t^{\prime}=1}^{n^{\prime}}\sum_{s=1}^{S}\int_{v\in \mathcal{I}}\big[\X_{t^{\prime}s}(v) - \widehat{\X}_{t^{\prime}s}(v)\big]^2dv,\label{eq:MSFE}
\end{align}
where $t^{\prime}$ is the index of the testing data for a total of $n^{\prime}$ observations and $\widehat{\X}_{ts}(v)$ is the prediction by replacing the true coefficient surfaces with the estimated ones as in~\eqref{eq:9}. To evaluate the predictive and estimation performance of the proposed method, we consider three competing approaches: 
\begin{inparaenum}
\item[(1)] no sparsity penalty;
\item[(2)] a global penalty, where the penalization is applied uniformly across the entire support of the coefficient surfaces; and 
\item[(3)] a combined global/local penalty, which applies both global penalization and additional localized penalization over specific regions. 
\end{inparaenum}

\subsection{Simulation results}

The comparative prediction results for these scenarios are summarized in Table~\ref{tab:simulationprediction}. We will use \textit{FBM} to denote our model with bivariate splines expressing the coefficient surface. \textit{FBM: Global penalty} represents only using a global sparsity penalty, where as \textit{FBM: Global/Local penalty} uses the two-level penalty as in~(\ref{eq:grouplasso}).
\begin{table}[!htb]
\tabcolsep 0.22in
\centering
\caption{The MAFE and MSFE of different settings in forecasting HDFTS for different lengths of time series, i.e., $n = 50, 100, 200, 500$. The means of MAFE and MSFE for the $1{,}000$ simulation are reported along with their standard deviation values in the parentheses.}
\label{tab:simulationprediction}
\begin{tabular}{@{}lccc@{}}
\toprule
   &  & MAFE &  \\
\toprule
$T$     & FBM: No sparse penalty & FBM: Global penalty &  FBM: Global/Local penalty \\
\midrule
50  &  0.284(0.026) & 0.127(0.008)  & 0.120(0.007) \\
100 &  0.280(0.032) & 0.126(0.009)  & 0.119(0.008) \\
200 &  0.287(0.025) & 0.129(0.008)  & 0.121(0.009) \\
500 &  0.280(0.037) & 0.126(0.009)  & 0.118(0.008) \\
\toprule
  &  & MSFE &  \\
\toprule
$T$     & FBM: No sparse penalty & FBM: Global penalty &  FBM: Global/Local penalty \\
\midrule
50  &  0.138(0.025) & 0.027(0.004)  & 0.024(0.003) \\
100 &  0.135(0.027) & 0.027(0.004)  & 0.023(0.004) \\
200 &  0.142(0.026) & 0.028(0.004)  & 0.025(0.005) \\
500 &  0.137(0.035) & 0.027(0.004)  & 0.022(0.003) \\
\bottomrule
\end{tabular}
\end{table}

In terms of forecasting target functional time series, the proposed model incorporating global/local penalties demonstrates approximately a 20\% improvement in both MAFE and MSFE compared to the model without any sparsity penalty. However, the performance difference between the global and global/local penalization strategies is relatively modest in the simulation setting.

To further assess the estimation performance, we compare the recovery of the self coefficient $\beta_{ss}$, the partially sparse coefficient surfaces, as illustrated in Figure~\ref{fig:coefsurfaces}. For this purpose, we use the integrated squared error (ISE), defined as follows:
\begin{equation*}
\text{ISE}(\beta_{sg}) = \int_{u,v\in \mathcal{I}}\big[\beta_{sg}(u,v) - \widehat{\beta}_{sg}(u,v)\big]^2dudv.
\end{equation*}
We report the integrated squared errors (ISEs) for three distinct types of coefficient surfaces, corresponding to the columns in Figure~\ref{fig:coefsurfaces}, including the self-coefficient $\beta_{ss}$. The comparative results are summarized in Table~\ref{tab:simulationestimation}.\begin{table}[!htb]
\centering
\tabcolsep 0.15in
\caption{The ISE of three kinds of coefficient surfaces for different length $n = 50, 100, 200, 500$. The mean of ISE for the $1,000$ simulation is reported along with its standard deviation in the parentheses. }
\label{tab:simulationestimation}
\begin{small}
\begin{tabular}{@{}llccc@{}}
\toprule
        &       & \multicolumn{3}{c}{FBM} \\
  Type  & $n$   & No penalty & Global penalty & Global/Local penalty \\
\midrule
Self coefficient 
        & 50    &  0.141 (0.058) & 0.135 (0.052) & \textbf{0.063} (0.017) \\
        & 100   &  0.203 (0.067) & 0.202 (0.065) & \textbf{0.116} (0.047) \\
        & 200   &  0.170 (0.030) & 0.161 (0.029) & \textbf{0.094} (0.031) \\
        & 500   &  0.224 (0.021) & 0.221 (0.020) & \textbf{0.106} (0.017) \\
\midrule
Shape I 
        & 50    &  0.131 (0.075) & 0.133 (0.072) & \textbf{0.060} (0.013) \\
        & 100   &  0.175 (0.051) & 0.174 (0.050) & \textbf{0.102} (0.044) \\
        & 200   &  0.165 (0.021) & 0.161 (0.020) & \textbf{0.094} (0.020) \\
        & 500   &  0.223 (0.017) & 0.222 (0.017) & \textbf{0.099} (0.010) \\
\midrule
Shape II 
        & 50    &  0.142 (0.064) & 0.140 (0.062) & \textbf{0.035} (0.019) \\
        & 100   &  0.200 (0.065) & 0.193 (0.064) & \textbf{0.103} (0.032) \\
        & 200   &  0.162 (0.024) & 0.153 (0.023) & \textbf{0.114} (0.000) \\
        & 500   &  0.224 (0.023) & 0.215 (0.022) & \textbf{0.091} (0.014) \\
\midrule
Shape III 
        & 50    &  0.146 (0.085) & 0.141 (0.083) & \textbf{0.041} (0.017) \\
        & 100   &  0.183 (0.067) & 0.182 (0.065) & \textbf{0.097} (0.036) \\
        & 200   &  0.174 (0.033) & 0.174 (0.032) & \textbf{0.084} (0.025) \\
        & 500   &  0.222 (0.041) & 0.213 (0.040) & \textbf{0.093} (0.030) \\
\bottomrule
\end{tabular}
\end{small}
\end{table}

The simulation results demonstrate that the proposed method that incorporates both global and local sparsity controls outperforms alternatives that either lack sparsity or employ only global sparsity in estimating the coefficient surfaces. Notably, the method with only global sparsity does not yield substantial improvements over the non-sparse model. When the sparsity parameter is selected using a validation set, the global-only approach often struggles to distinguish between coefficient surfaces, occasionally selecting a minimal $\lambda$ that fails to adequately shrink the three non-significant surfaces affected by random noise. In general, simulation studies confirm that the proposed method with combined global and local penalties achieves superior performance in both estimation accuracy and forecasting.

Table~\ref{tab:degree_sensitivity_wide} presents the sensitivity analysis results for the choice of basis degree $d$ in the bivariate spline basis functions. The results indicate that the model's performance is not highly sensitive to the choice of $d$ as long as it is sufficiently large.
\begin{table}[!htb]
\centering
\caption{Sensitivity to basis degree $d$: MSFE (mean $\pm$ std) and ISE of estimated coefficient surfaces (mean $\pm$ std). }
\label{tab:degree_sensitivity_wide}
\small
\begin{tabular}{@{}lccccc@{}}
\toprule
 & \multicolumn{5}{c}{Degree $d$} \\
\cmidrule(lr){2-6}
Measure & 1 & 2 & 3 & 4 & 5 \\
\midrule
MSFE  
& $0.395 \pm 0.068$ 
& $0.312 \pm 0.052$ 
& $0.300 \pm 0.057$ 
& $0.298 \pm 0.049$ 
& \textbf{$0.288 \pm 0.060$} \\
\addlinespace
ISE (Self) 
& \textbf{$0.017 \pm 0.008$} 
& $0.026 \pm 0.011$ 
& $0.034 \pm 0.014$ 
& $0.043 \pm 0.015$ 
& $0.046 \pm 0.017$ \\
ISE (Shape I) 
& \textbf{$0.301 \pm 0.090$} 
& $0.313 \pm 0.094$ 
& $0.327 \pm 0.098$ 
& $0.339 \pm 0.101$ 
& $0.350 \pm 0.102$ \\
ISE (Shape II) 
& \textbf{$0.264 \pm 0.070$} 
& $0.285 \pm 0.072$ 
& $0.301 \pm 0.078$ 
& $0.317 \pm 0.079$ 
& $0.331 \pm 0.081$ \\
ISE (Shape III) 
& $0.147 \pm 0.022$ 
& $0.149 \pm 0.020$ 
& $0.146 \pm 0.021$ 
& $0.148 \pm 0.022$ 
& \textbf{$0.140 \pm 0.020$} \\
\bottomrule
\end{tabular}
\end{table}

\begin{figure}[!htb]
\includegraphics[width=\textwidth]{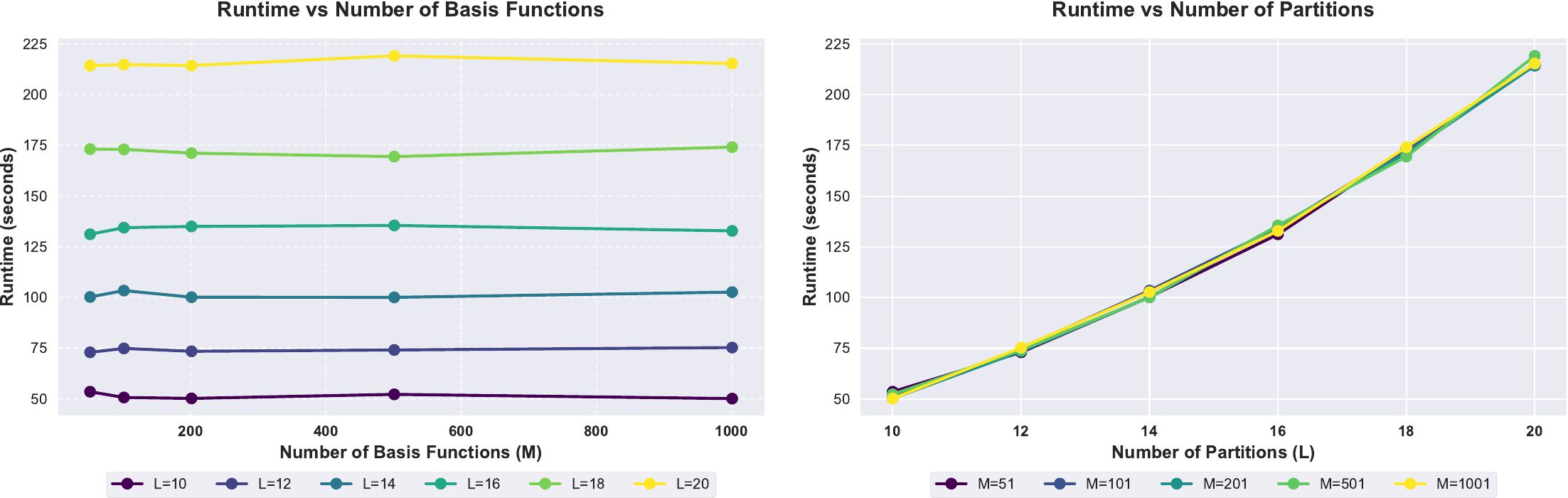}
\caption{\label{fig:runtime} The computation time (in seconds) for different combinations of $M$ and $L$. Each cell represents the average computation time over 10 runs, each with a pseudo-random seed.}
\end{figure}

From Figure~\ref{fig:runtime}, we can see that the computation time does not increase significantly as $M$ increases. However, the computation time increases more noticeably as $L$ increases. The reason is that the design matrix (after evaluating the basis functions at $M$ points and considering a single predictor at a time) is a block matrix with $L$ blocks, and where each block has $Q$ columns. Each block corresponds to one triangle in the triangulation, and the number of rows is determined by the number of points in that triangle. Hence, the computation time, which will mostly be an iterative solution, is more sensitive to $L$ (in relation to $Q$) than $M$.

\section{Japanese subnational age-specific mortality rates}\label{sec:4}

Mortality rate serves as a fundamental indicator of a nation's health status, quantifying the number of deaths within a specific population at a given age. Rather than analyzing national-level mortality data alone, we consider mortality rates across various regions, yielding a richer understanding of regional health dynamics. However, this regional granularity introduces additional complexities in statistical modeling and forecasting. Beyond predicting regional mortality trends, our interest also lies in uncovering inter-regional relationships and assessing how mortality trends in one region may influence those in another.

\begin{figure}[!htb]
\centering
\includegraphics[width=\textwidth]{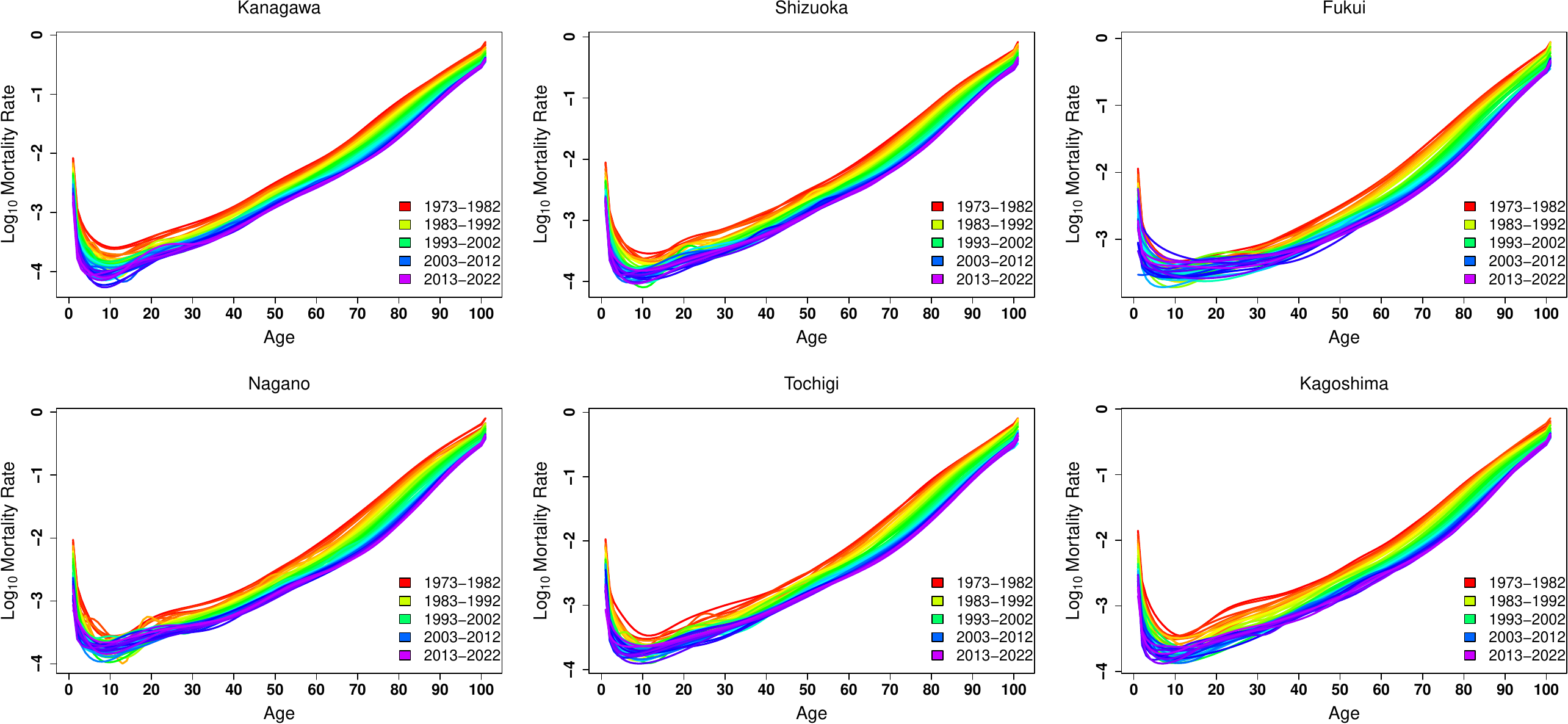}
\caption{Smoothed $\log_{10}$ mortality rate curves from 1973 to 2022 for six randomly chosen prefectures (with their names as the title) in Japan. The color represents the year of the mortality rate curve, from red (oldest) to purple (most current).}\label{fig:data_preview}
\end{figure}

In this study, we apply the proposed methodology to investigate the temporal evolution of mortality across sub-regions of a country. As a case study, we analyze Japan, which comprises 47 prefectures, for which age-specific mortality data are available from 1973 to 2022. Treating age-specific mortality rates as functional data, the collection of curves across prefectures constitutes HDFTS. Figure~\ref{fig:data_preview} presents a preview of the mortality rate curves for six randomly selected prefectures in Japan, illustrating the variability and structure inherent in the data.

Since mortality rates at older ages may exceed one due to small population denominators, we apply a nonparametric smoothing technique to stabilize and regularize the data. Specifically, we implement monotonically constrained penalized regression splines to ensure smooth and biologically plausible mortality trajectories. Let $N_t(u)$ denote the total population of age $u$ on June~30 in year $t$. Assuming binomial variability, the observed mortality rate $m_t(u)$ is approximately distributed as a binomial proportion with estimated variance $N_t^{-1}(u) m_t(u)[1 - m_t(u)]$. Applying a first-order Taylor approximation, the variance of the log-mortality rate $\log[m_t(u)]$ is approximately given by:
\begin{equation*}
\widehat{\sigma}^2_t(u)\approx [1-m_t(u)]N_t^{-1}(u)m_t^{-1}(u).
\end{equation*}
We define weights equal to the inverse variances $w_t(u) = N_t(u)m_t(u)/[1-m_t(u)]$ and use weighted penalized regression splines in \cite{Wood03} and \cite{HN99} to estimate the curve in each year.

Figure~\ref{fig:data_preview} illustrates a general decreasing trend in mortality rate curves over time, reflecting improvements in population health across all prefectures in Japan. This temporal pattern is consistently observed, although regional differences remain evident. For example, mortality rates in Iwate appear persistently higher than in other prefectures, and the temporal decline is less pronounced. A similar pattern is observed in Akita, where the trend is comparatively less clear. In contrast, prefectures such as Aomori and Yamagata exhibit a notable decrease in mortality rates for groups aged 8–16. These observed variations highlight potential spatial heterogeneity and motivate further investigation of the underlying patterns and interdependencies across regions. To do this, we apply the additive model in~\eqref{eq:1} to capture and analyze both temporal dynamics and spatial dependencies in the mortality rate curves across Japan’s prefectures.
\begin{figure}[!htb]
\includegraphics[width=\textwidth]{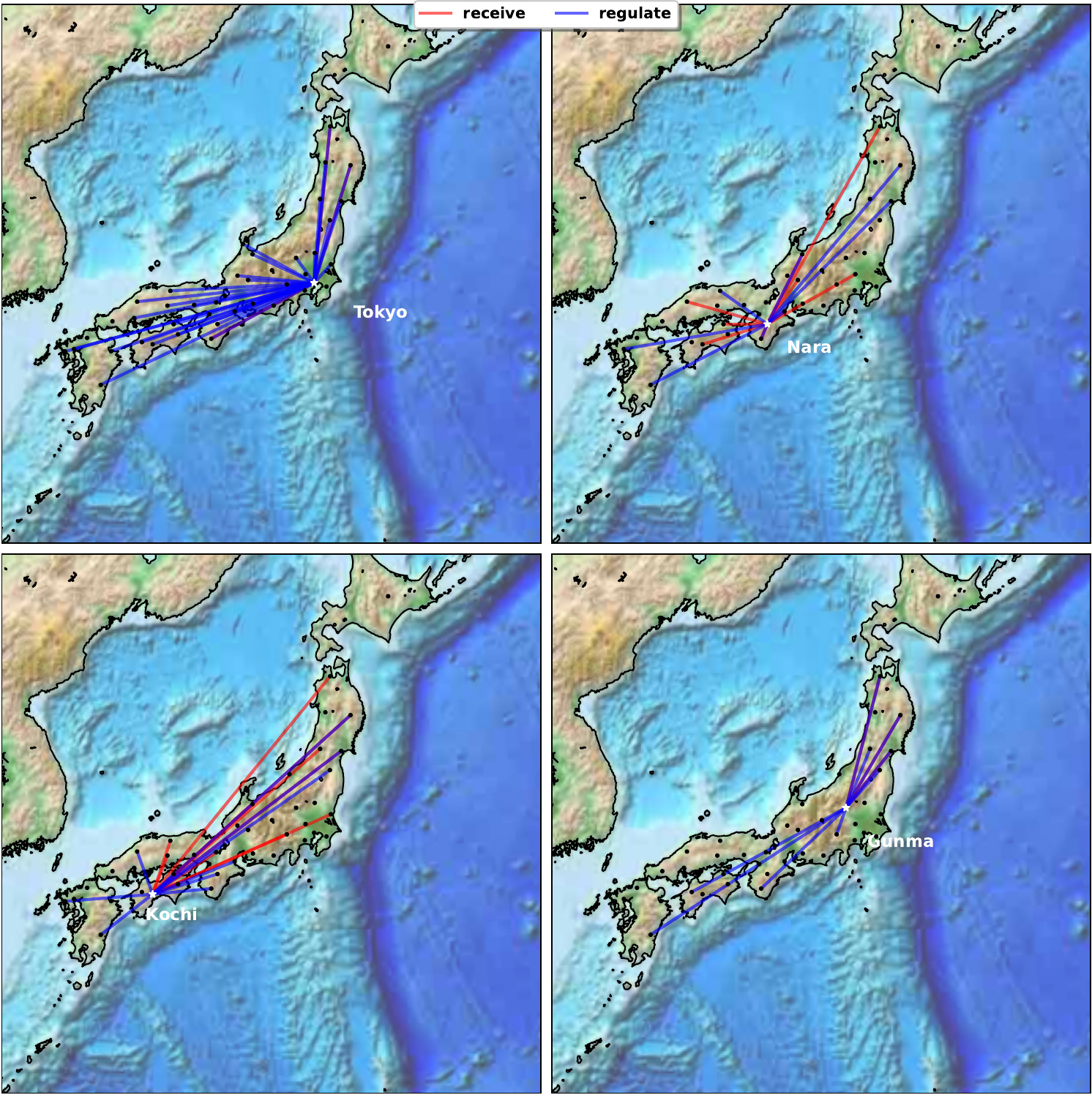}
\caption{The figure shows retrieved underlying connections between a target prefecture (indicated by the white star) and all others. The red line indicates a significant receiving connection from another to the target region, and the blue line indicates a significant regulating connection from the target region to another. }\label{fig:region_connection_map}
\end{figure}

To investigate the interdependencies in mortality rates among Japan’s prefectures, we begin by constructing a forecast model with $\delta = 1$, capturing one-period-ahead relationships. Figure~\ref{fig:region_connection_map} visualizes the inferred spatial connections for a given target prefecture, marked by a white star. The red lines represent significant receiving connections, indicating that the mortality trend in the target prefecture is influenced by another region, while the blue lines denote significant regulating connections, where the target prefecture influences mortality trends in other regions. For instance, in the case of Tokyo, which is the most populous and densely populated prefecture, the model identifies mostly the regulating connections to other regions. This suggests that Tokyo’s mortality dynamics significantly influences those of other prefectures. In contrast, the model detects only a few significant receiving connections to Tokyo, specifically from Aomori, Iwate, and Wakayama, indicating a relatively limited degree of external influence on Tokyo’s mortality trends.

\begin{figure}[!htb]
\centering
\includegraphics[width=18cm]{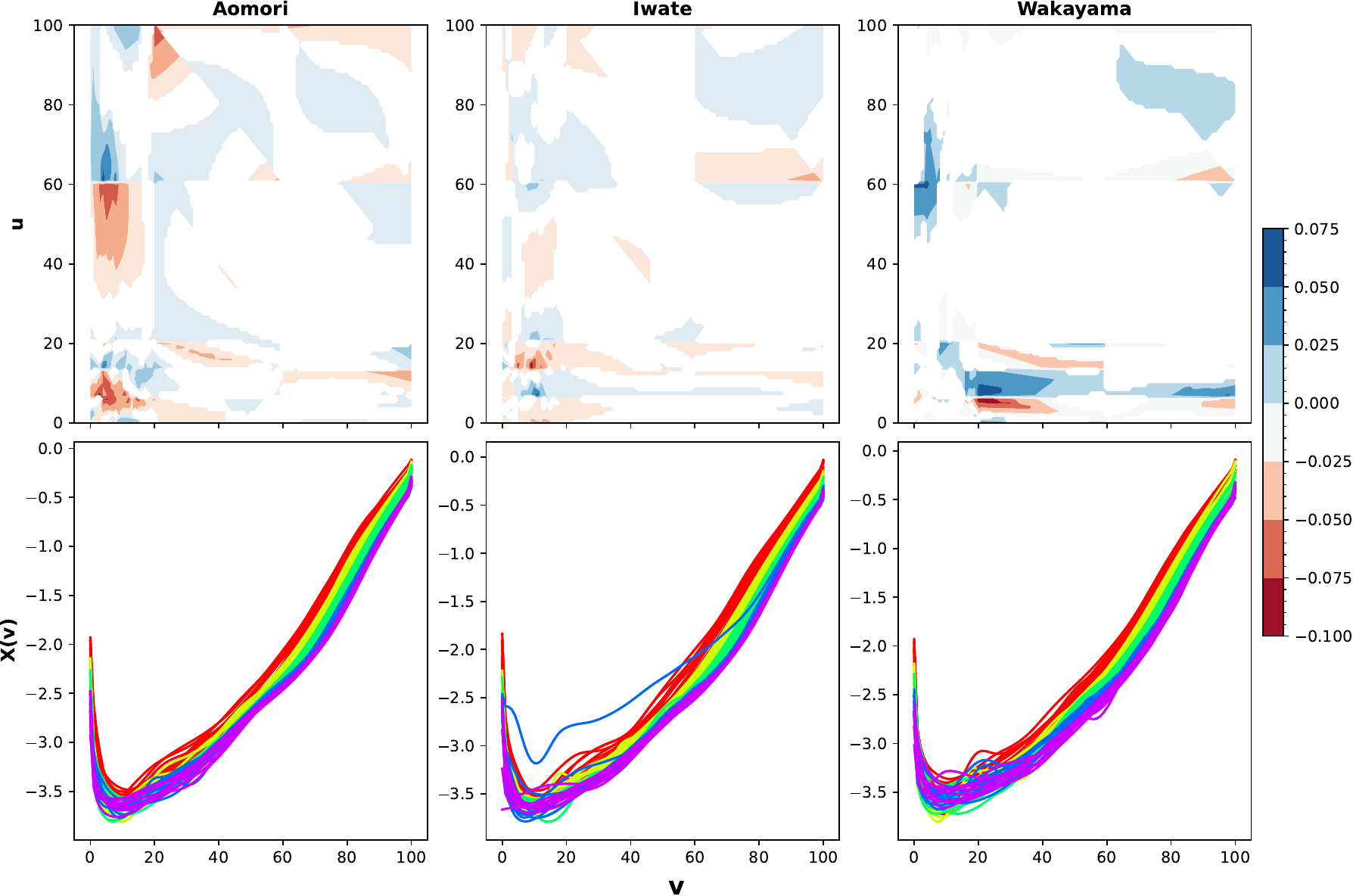}
\caption{The coefficient surfaces (in the top row) of prefectures (Aomori, Iwate, and Wakayama) affect Tokyo's mortality rate. The bottom row compares the mortality rate curves of three prefectures with the color scale identical to Figure~\ref{fig:data_preview}.To obtain the mortality rate at a specific age of the target region, we take the sum of inner products between a curve from the predictors and a column-wise slice of the corresponding coefficient surface. The $u, v$ denote the age range.}\label{fig:coefficient_Tokyo}
\end{figure}

We examine the estimated coefficient surfaces associated with the regions that influence Tokyo's mortality rate, as depicted in Figure~\ref{fig:coefficient_Tokyo}. The global sparsity structure facilitates the identification of significant spatial dependencies, highlighting which prefectures exert a notable influence on the mortality dynamics of Tokyo. In addition, the local sparsity mechanism offers finer resolution by revealing age-specific effects within these relationships. This dual sparsity structure not only enhances interpretability but also allows for a more nuanced understanding of how particular age groups contribute to the inter-regional dependencies in mortality patterns.

In the process of deriving the target mortality rate curve, we compute the inner product between the coefficient surfaces and the mortality rate curves of the regulatory prefectures. For a simple illustration, consider the model $\X_{s}(v) = \int_{\mathcal{I}} \beta(u,v) \X_{s^{\prime}}(u) du$. This is analogous to calculating the inner product between a curve in the bottom panel and the column slice (at a specific age) of the coefficient surface shown in the top panel of Figure~\ref{fig:coefficient_Tokyo}. By examining the coefficient surfaces, we gain insights into how a predictor influences the mortality rate of the target region at a specific age.

In general, the mortality rate curves of the target region exhibit greater variability with respect to predictors for the group aged 0-20. Notably, the coefficient surfaces of Aomori and Iwate show more similarity to each other than to those from Wakayama. Specifically, the mortality rate of Tokyo for the group aged 0-20 is influenced by the same age group and population group (ages 40-80) in Aomori. Similarly, the mortality rate of Tokyo for the age group 0-20 is also influenced by the same age group in Iwate. However, for the group aged 20-100+, the mortality rate in Tokyo is more strongly influenced by the Wakayama mortality rate for the group aged 0-20, compared to the other two prefectures. This highlights the spatial and age-specific dependencies in mortality patterns between regions.

The MAFE and MSFE of the prediction are summarized in Table~\ref{tab:1}, where the MAFE and MSFE are defined in~\eqref{eq:MAFE} and~\eqref{eq:MSFE}. We assess the performance of the proposed model in forecasting mortality rate curves in Japan across different forecast horizons or time lags, denoted by $\delta$. 
For comparison purposes, we consider several existing methods: univariate functional time series (UFTS) \citep[see, e.g.,][]{HS09}; multivariate functional time series (MFTS) \citep[see, e.g.,][]{SK22}; multilevel functional time series (MLFTS) \citep[see, e.g.,][]{TSY22}; and nonlinear prediction of functional time series (NOP) \citep[see, e.g.,][]{NOP}. In the UFTS method, a time series of functions is modeled by a set of functional principal components and their associated scores, extracted from the covariance of the functions. The scores are extrapolated with the univariate time-series forecasting method. The scores' forecasts are then multiplied by the estimated functional principal components to obtain the forecast curves. In the MFTS method, multiple time series of functions are first stacked in a matrix of long vector. From its estimated covariance, we extract the multivariate functional principle components and their corresponding scores. In the MLFTS method, multiple time series of functions are summarized by a series-specific mean, common trend shared by different populations, series-specific residual trend, and error term. Through two analyses of the functional principal components, the common trend (i.e., the average of all series) and the residual trend are modeled and forecasted.
\begin{table}[!htb]
\centering
\tabcolsep 0.2in
\caption{\label{tab:1} The prediction errors (MAFE and MSFE) of different methods in forecasting the mortality rate curves in testing data for $\delta = 1, 5, 10$ and averaged over all prefectures. Nonlinear prediction of functional time series (NOP). Univariate functional time series (UFTS), Multivariate functional time series (MFTS), and Multilevel functional time series (MLFTS).}
\begin{tabular}{@{}llcccccc@{}}
\toprule
Lag &Error   & FBM & FBM with refit & NOP &  UFTS & MFTS & MLFTS \\
\midrule
$\delta = 1$ & MAFE & 0.042 & 0.039 & 0.045 & 0.042 & 0.045 & 0.046 \\
             & MSFE & 0.053 & 0.051 & 0.056 & 0.053 & 0.058 & 0.054 \\
\bottomrule
\midrule
$\delta = 5$ & MAFE & 0.053 & 0.047 & 0.055 & 0.053 & 0.060 & 0.054 \\
            & MSFE & 0.063 & 0.060 & 0.065 & 0.065 & 0.074 & 0.066 \\
\bottomrule

\midrule
$\delta = 10$ & MAFE & 0.059 & 0.057& 0.055 & 0.085 & 0.089 & 0.085 \\
            & MSFE & 0.059 & 0.059 &0.055 & 0.085 & 0.090 & 0.085 \\
\bottomrule
\end{tabular}
\end{table}

\begin{figure}[!htb]
\centering
\includegraphics[width=14.3cm]
{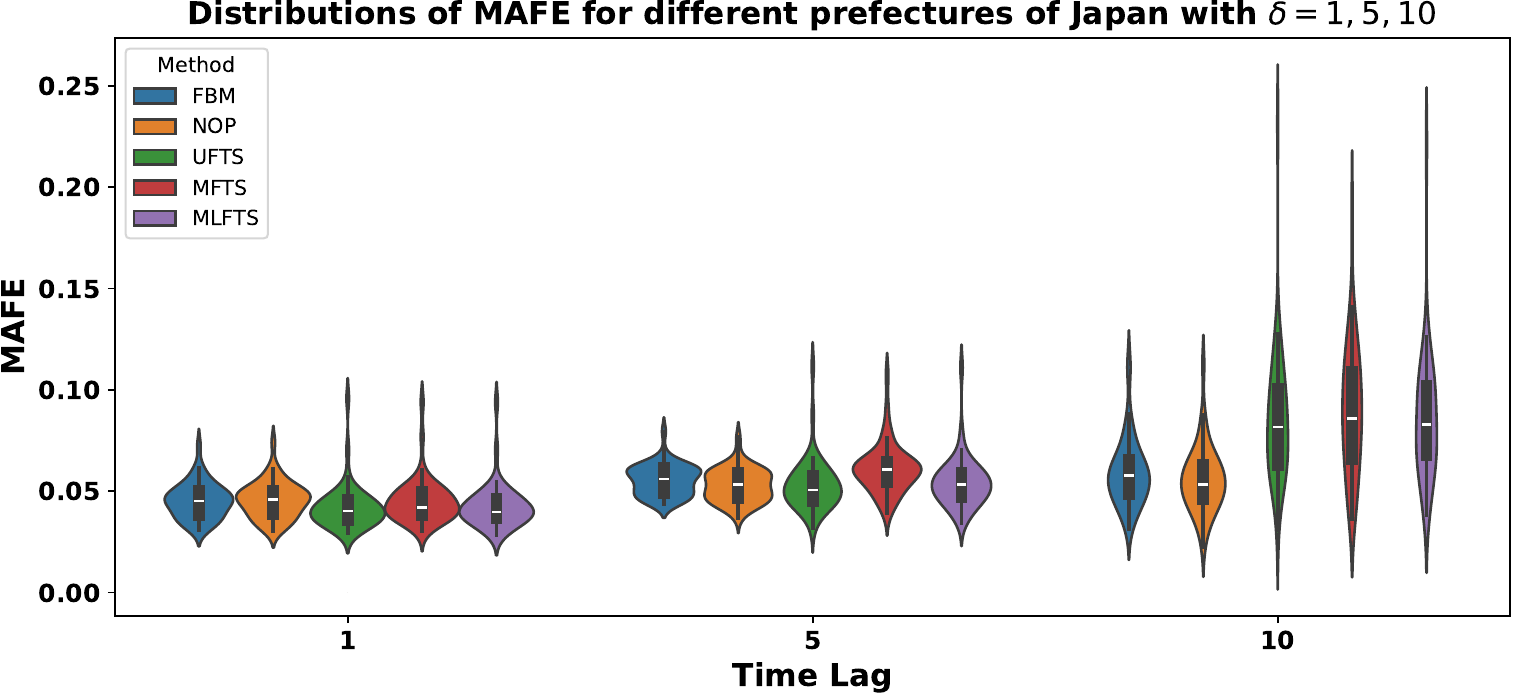}
\caption{The violin plot of MAFE for all prefectures in Japan for all methods at a time lag $\delta = 1, 5, 10$.}\label{fig:MAFEacrossprefectures}
\end{figure}

The proposed model demonstrates performance comparable to that of existing methods in forecasting mortality rate curves in Japan. Additionally, its forecast performance exhibits greater stability across different time lags, $\delta$, compared to other linear methods. While the long-term forecast performance of the proposed model slightly trails behind that of the machine learning method NOP, it offers significantly more interpretable results. Furthermore, the distribution of MAFEs across all prefectures in Japan for each method with time lags $\delta = 1, 5, 10$ is presented in Figure~\ref{fig:MAFEacrossprefectures}, illustrating the stability of the proposed model in producing reliable long-term forecast curves.

\section{Conclusion}\label{sec:5}

In this paper, we introduce an interpretable additive model for analyzing high-dimensional functional time series. Our model effectively captures the relationships between functional predictors and responses across different regions and time points, providing valuable information on the complex interactions within the HDFTS. By incorporating local sparse estimation and penalized smoothing bivariate splines, our approach not only enhances predictive performance but also provides interpretability. Through simulation studies and empirical applications to age-specific mortality rates in Japan, we demonstrated that our model improves prediction accuracy while offering superior interpretability. In particular, the empirical application revealed the model's ability to identify significant age-specific regions of the coefficient surfaces, thereby improving our understanding of how mortality rates have evolved across various regions of Japan.

Despite the strengths of our approach, several challenges remain. One limitation observed in the application of the mortality rate is the tendency for mortality rates from other regions to primarily influence the current region and age when the age groups are closely related. This pattern could be captured more effectively by incorporating structured groups, such as nested groups, within the sparsity penalty. Furthermore, an important avenue for future work involves expanding the model to account for more complex dependencies, including spatio-temporal interactions and multivariate HDFTS. Such extensions would enhance the model's applicability to a wider range of problems, such as climate modeling and financial forecasting, where capturing intricate, dynamic relationships is crucial.

\section*{Supplementary Materials}

\begin{description}
\item[Online Shiny app] We have provided a Shiny app to examine the predictive performances for each prefecture in Japan \url{https://haixuw.shinyapps.io/FBM-HDFTS/}
\item[\Rlogo\ codes] We have the reproducible codes available for running the proposed FBM model on Japan's mortality rate curves \url{https://github.com/alex-haixuw/FBM-HDFTS/}
\end{description}

\section*{\large Appendix}

\noindent\textit{\textbf{Proof of Theorem~\ref{thm:1}}}:
\begin{proof}
First, we introduce an estimator $\widehat{\bm{\gamma}}(0,\lambda_{2})$ that is the minimizer of the objective function~\ref{eq:6} with the setting $\lambda_{1} = 0$. By the definition of $\widehat{\bm{\gamma}}$, we have the following:
\begin{align*}
  \mathcal{L}_{n}(\widehat{\bm{\gamma}}) &\leq \mathcal{L}_{n}(\widehat{\bm{\gamma}}(0,\lambda_{2}))\\
  \mathcal{P}_{1}(\gammanosparse; \lambda_{1}) - \mathcal{P}_{1}(\gammaest; \lambda_{1}) &\geq ||\bm{y} - \bm{\Psi}\gammaest||^{2}_{2} - ||\bm{y} - \bm{\Psi}\gammanosparse||^{2}_{2} + \mathcal{P}_{2}(\gammaest; \lambda_{2}) - \mathcal{P}_{2}(\gammanosparse; \lambda_{2})
\end{align*}
We will omit $(\cdot; \lambda)$ to $(\cdot)$ for simplicity, although the penalty parameter $\lambda_{1}$ is still in the term. For global selection, we observe that a single penalty on $c_{g}||\bm{\gamma}_{g}||^{\nu}_{1} = c_{g,L+1}||\bm{\gamma}_{g,L+1}||_{1}^{\nu}$ is enough. We will reduce the sum $\sum_{l=1}^{L+1}c_{g,l}||\bm{\gamma}_{g,l}||^{\nu}_{1}$ to this single penalty for the $g$\textsuperscript{th} coefficient surface. The difference on the left of the inequality can be written as:
\begin{align*}
  \mathcal{P}_{1}(\gammanosparse) - \mathcal{P}_{1}(\gammaest) &= \lambda_{1} \sum_{g=1}^{S}c_{g} \Big\{ ||\gammanosparseG||^{\nu}_{1} - ||\gammaestG||^{\nu}_{1} \Big\} \\
  &\leq 2\lambda_{1}\sum_{g=1}^{S}c_{g} \Big\{ ||\gammanosparseG||^{\nu - 1}_{1}||\gammanosparseG - \gammaestG||_{1}\Big\} \\
  &\leq 2\lambda_{1}\sum_{g=1}^{S}c_{g} ||\gammanosparseG||^{\nu - 1}_{1} \Big\{ c_{g}||\gammanosparseG - \gammaestG||^{2}_{2} \big\}^{\frac{1}{2}} \\
  &\leq 2\lambda_{1}(\sum_{g=1}^{S}c_{g}^{3}||\gammanosparseG||^{2\nu-2}_{1})^{\frac{1}{2}}(\sum_{g=1}^{S}||\gammanosparseG - \gammaestG||^{2}_{2})^{\frac{1}{2}}
\end{align*}
by Cauchy-Schwarz inequality. Furthermore, we can see that 
\begin{equation*}
  \sum_{g=1}^{S}||\gammanosparseG - \gammaestG||^{2}_{2} \leq ||\gammanosparse - \gammaest||^{2}_{2},
\end{equation*}
since there is no overlapping in coefficients across different $g = 1,..., S$. The lower bound of $\mathcal{P}_{1}(\gammanosparse) - \mathcal{P}_{1}(\gammaest)$ is the sum of differences between quadratic terms, i.e., 
\begin{align*}
  &||\bm{y} - \bm{\Psi}\gammaest||^{2}_{2} - ||\bm{y} - \bm{\Psi}\gammanosparse||^{2}_{2} + \mathcal{P}_{2}(\gammaest; \lambda_{2}) - \mathcal{P}_{2}(\gammanosparse; \lambda_{2})\\
  &= (\bm{y} - \bm{\Psi}\gammaest)^{\top}(\bm{y} - \bm{\Psi}\gammaest) - (\bm{y} - \bm{\Psi}\gammanosparse)^{\top}(\bm{y} - \bm{\Psi}\gammanosparse) + \lambda_{2}\gammaest^{\top}\bm{R}\gammaest - \lambda_{2}\gammanosparse^{\top}\bm{R}\gammanosparse\\
  &= \widehat{\bm{\Delta}}^{\top}(\bm{\Psi}^{\top}\bm{\Psi} +\lambda_{2}\bm{R})\widehat{\bm{\Delta}} \quad\text{where}\quad \widehat{\bm{\Delta}} = \gammaest - \gammanosparse\\
  &\geq (\rho_{\min}(\bm{\Psi}^{\top}\bm{\Psi}  + \lambda_{2}\bm{R}))||\widehat{\bm{\Delta}}||^{2}_{2} \quad\text{where} \hspace*{0.5em} \rho_{\min}(\cdot) \text{ is the minimum eigenvalue of the matrix.}
\end{align*}
Combining the two inequalities on $\mathcal{P}_{1}(\gammanosparse) - \mathcal{P}_{1}(\gammaest)$, we have
\begin{align*}
\rho_{\min}||\widehat{\bm{\Delta}}||^{2}_{2} &\leq 2\lambda_{1}(\sum_{g=1}^{S}c_{g}^{3}||\gammanosparseG||^{2\nu-2}_{1})^{\frac{1}{2}}(\sum_{g=1}^{S}||\gammanosparseG - \gammaestG||^{2}_{2})^{\frac{1}{2}}\\
||\widehat{\bm{\Delta}}||^{2}_{2} &\leq \frac{2\lambda_{1}}{\rho_{\min}}(\sum_{g=1}^{S}c_{g}^{3}||\gammanosparseG||^{2\nu-2}_{1})^{\frac{1}{2}} = \frac{2\lambda_{1}\eta_{\gamma}}{\rho_{\min}}
\end{align*}
Before bounding our estimator $\gammaest$ with the true value $\gammatrue$, we need one more step which is to bound $||\gammanosparse - \gammatrue||^{2}_{2}$. That is, we first define 
\begin{equation*}
  \widehat{\bm{\Delta}}_{\star} = \gammanosparse - \gammatrue = -\lambda_{2}(\bm{\Psi}^{\top}\bm{\Psi} + \lambda_{2}\bm{R})^{-1}\gammatrue + (\bm{\Psi}^{\top}\bm{\Psi} + \lambda_{2}\bm{R})^{-1}\bm{\Psi}^{\top}\bm{\epsilon}
\end{equation*}
which is the difference between the Ridge-like estimator and true coefficients. Then, we can derive the expected square of the difference as follows:
\begin{align*}
  \mathbb{E}[||\widehat{\bm{\Delta}}_{\star}||^{2}_{2}] &\leq 2\mathbb{E}[||\lambda_{2}(\bm{\Psi}^{\top}\bm{\Psi} + \lambda_{2})^{-1}\gammatrue||^{2}_{2}] + 2\mathbb{E}[||(\bm{\Psi}^{\top}\bm{\Psi} + \lambda_{2})^{-1}\bm{\Psi}^{\top}\bm{\epsilon}||^{2}_{2}]\\
  &\leq 2\lambda_{2}^{2}\rho_{\min}^{-2}||\gammatrue||^{2}_{2} + \rho_{\min}^{-2}\mathbb{E}[||\bm{\Psi}^{\top}\bm{\epsilon}||^{2}_{2}]\\
  &\leq 2\rho_{\min}^{-2}(\lambda_{2}^{2}||\gammatrue||^{2}_{2} + \mathbb{E}[||\bm{\Psi}\bm{\epsilon}||^{2}_{2}])\\
  &\leq 2\rho_{\min}^{-2}(\lambda_{2}^{2}||\gammatrue||^{2}_{2} + nKM\rho_{\max}\sigma^{2})
\end{align*}
Now, we are able to show that the estimator is convergent to the true value in mean squared error sense. We have
\begin{align*}
\mathbb{E}(||\gammaest - \gammatrue||^{2}_{2}) &= \mathbb{E}(||\widehat{\bm{\Delta}} + \widehat{\bm{\Delta}}_{\star}||^{2}_{2})\\ 
&\leq 2\mathbb{E}(||\widehat{\bm{\Delta}}||^{2}_{2}) + 2\mathbb{E}(||\widehat{\bm{\Delta}}_{\star}||^{2}_{2})\\
&\leq \frac{4\lambda_{1}\eta_{\gamma}}{\rho_{\min}^{2}} + 4\rho_{\min}^{-2}(\lambda_{2}^{2}||\gammatrue||^{2}_{2} + nK\rho_{\max}\sigma^{2})\\
&\leq \frac{4\lambda_{1}\eta_{\gamma} + 4(\lambda_{2}^{2}||\gammatrue||^{2}_{2} + nK\rho_{\max}\sigma^{2})}{\rho_{\min}^2} \\
&\leq  4\frac{\lambda^{2}_{1}\eta^{2}_{\gamma} + \lambda_{2}^{2}||\gammatrue||_{2}^{2} + nKb\sigma^{2}}{(nKa + \lambda_{2})^{2}}
\end{align*}
\end{proof}

\noindent\textit{\textbf{Proof of Theorem~\ref{thm:2}}}:
\begin{proof}
We observe that the optimization of the objective function~\eqref{eq:6} satisfies the Karush-Kuhn-Tucker (KKT) conditions. The KKT conditions imply that the solution $\widehat{\bm{\gamma}}$ satisfies the following conditions:
\begin{equation*}
2(\bm{y} - \bm{\Psi}\widehat{\bm{\gamma}})^{\top}\bm{\Psi}_{glq} - \lambda_{2}\widehat{\bm{\gamma}}_{glq} =  \lambda_{1}\nu c_{g}||\gammaestG||^{\nu-1}_{1}sgn(\gammaest_{glq})
\end{equation*}
for $g \in B_{1}$, $l = 1,..., L$ and $q = 1,..., Q$, where $\bm{\Psi}_{glq}$ is the corresponding column in $\bm{\Psi}$. In the meantime,  for any coefficient in the non-significant $g \in B_{2}$, we have
\begin{equation*}
2(\bm{y} - \bm{\Psi}\widehat{\bm{\gamma}})^{\top}\bm{\Psi}_{glq} - \lambda_{2}\widehat{\bm{\gamma}}_{glq} < \lambda_{1}\nu c_{g}||\gammaestG||^{\nu-1}_{1}sgn(\gammaest_{glq}).
\end{equation*}
To prove the selection consistency, it is sufficient to show that 
\begin{equation*}
P(\forall g \in B_{2}, |2(\bm{y} - \bm{\Psi}\widehat{\bm{\gamma}})^{\top}\bm{\Psi}_{glq} - \lambda_{2}\widehat{\bm{\gamma}}_{glq}| < \lambda_{1}\nu c_{g}||\gammaestG||^{\nu-1}_{1}sgn(\gammaest_{glq})) \to 1
\end{equation*}
First, we define the estimator $\widetilde{\bm{\gamma}}$ knowing the true set $B_{1}$ and $B_{2}$. That is, $\widetilde{\bm{\gamma}}_{g} = \gammaestG$ for $g \in B_{1}$ and $0$ for $g \in B_{2}$. Again, we use the definition of minimizer to show that 
\begin{align*}
  \mathcal{L}_{n}(\widehat{\bm{\gamma}}) &\leq \mathcal{L}_{n}(\widetilde{\bm{\gamma}}) \\
  \mathcal{P}_{1}(\widehat{\bm{\gamma}}) - \mathcal{P}_{1}(\widetilde{\bm{\gamma}}) & \leq ||\bm{y} - \bm{\Psi}\widetilde{\bm{\gamma}}||^{2}_{2} - ||\bm{y} - \bm{\Psi}\widehat{\bm{\gamma}}||^{2}_{2} + \mathcal{P}_{2}(\widetilde{\bm{\gamma}}) - \mathcal{P}_{2}(\widehat{\bm{\gamma}}) \\
  \lambda_{1}\sum_{g=1}^{S}c_{g}(||\widehat{\bm{\gamma}}_{g}||^{\nu}_{1} - ||\widetilde{\bm{\gamma}}_{g}||^{\nu}_{1}) &\leq ||\bm{\Psi}(\widehat{\bm{\gamma}} - \widetilde{\bm{\gamma}})||^{2}_{2} + \lambda_{2}(||\widehat{\bm{\gamma}}||_{2}^{2} - ||\widetilde{\bm{\gamma}}||_{2}^{2}) + 2(\bm{y} - \bm{\Psi}\widehat{\bm{\gamma}})^{\top}\bm{\Psi}(\widehat{\bm{\gamma}} - \widetilde{\bm{\gamma}})
\end{align*}
We will work on the left side of the inequality, and the KKT condition implies that
\begin{align*}
  2(\bm{y} - \bm{\Psi}\gammaest)^{\top}\bm{\Psi}(\gammaest - \widetilde{\bm{\gamma}}) &= \sum_{g \in B_{2}}||\gammaestG||_{1}\lambda_{1}\nu c_{g}||\gammaestG||^{\nu-1}_{1} \\
  &= \lambda_{1}\nu\sum_{g \in B_{2}}c_{g}||\gammaestG||^{\nu-1}_{1}(||\gammaestG||_{1} - ||\widetilde{\bm{\gamma}}_{g}||_{1})\\
  &=\lambda_{1}\nu\sum_{g=1}^{S}c_{g}||\gammaestG||^{\nu-1}_{1}(||\gammaestG||_{1} - ||\widetilde{\bm{\gamma}}_{g}||_{1}) \\
  &\leq \lambda_{1}\sum_{g \in B_{1}} c_{g}(||\gammaestG||^{\nu}_{1} - ||\widetilde{\bm{\gamma}}_{g}||^{\nu}_{1}) + \lambda_{1}\nu\sum_{g \in B_{2}}c_{g}||\widehat{\bm{\gamma}}_{g}||^{\nu}_{1}
\end{align*}
First, we start with a bit of rearranging of the above inequality. We have
\begin{align*}
  2(\bm{y} - \bm{\Psi}\gammaest)^{\top}\bm{\Psi}(\gammaest - \widetilde{\bm{\gamma}}) &+ \lambda_{1}(1 - \nu)\sum_{g \in B_{2}}c_{g}||\widehat{\bm{\gamma}}_{g}||^{\nu}_{1} \leq \lambda_{1}\sum_{g \in B_{1}} c_{g}(||\gammaestG||^{\nu}_{1} - ||\widetilde{\bm{\gamma}}||^{\nu}_{1}) - \lambda_{1}\sum_{g \in B_{2}} c_{g}||\gammaestG||^{\nu}_{1}\\
  &\leq \lambda_{1}\sum_{g=1}^{S} c_{g}(||\gammaestG||^{\nu}_{1} - ||\widetilde{\bm{\gamma}}||^{\nu}_{1})\\
  &= \mathcal{P}_{1}(\widehat{\bm{\gamma}}) - \mathcal{P}_{1}(\widetilde{\bm{\gamma}})\\
  &\leq  ||\bm{\Psi}(\widehat{\bm{\gamma}} - \widetilde{\bm{\gamma}})||^{2}_{2} + \lambda_{2}(||\widehat{\bm{\gamma}}||_{2}^{2} - ||\widetilde{\bm{\gamma}}||_{2}^{2}) + 2(\bm{y} - \bm{\Psi}\widehat{\bm{\gamma}})^{\top}\bm{\Psi}(\widehat{\bm{\gamma}} - \widetilde{\bm{\gamma}}),
\end{align*}
then
\begin{align*}
  \lambda_{1}(1 -\nu)\sum_{g \in B_{2}}c_{g}||\widehat{\bm{\gamma}}_{g}||^{\nu}_{1} &\leq ||\bm{\Psi}(\widehat{\bm{\gamma}} - \widetilde{\bm{\gamma}})||^{2}_{2} + \lambda_{2}(||\widehat{\bm{\gamma}}||_{2}^{2} - ||\widetilde{\bm{\gamma}}||_{2}^{2})\\
  \lambda_{1}(1-\nu)\sum_{g \in B_{2}}c_{g}||\widehat{\bm{\gamma}}_{g}||^{\nu}_{1} &\leq \rho_{\max}\sum_{g \in B_{2}}||\widehat{\bm{\gamma}}_{g}||_{2}^{2}\\
  c_{\max}\lambda_{1}(1-\nu)\sum_{g \in B_{2}}||\widehat{\bm{\gamma}}_{g}||^{\nu}_{1} &\leq \rho_{\max}\sum_{g \in B_{2}}||\widehat{\bm{\gamma}}_{g}||_{2}^{2}\\
  c_{\max}\lambda_{1}(1-\nu)||\widehat{\bm{\gamma}}_{g \in B_{2}}||^{\nu}_{2} &\leq \rho_{\max}||\widehat{\bm{\gamma}}_{g \in B_{2}}||^{2}_{2}
\end{align*}
where $\widehat{\bm{\gamma}}_{g \in B_{2}} = (\widehat{\bm{\gamma}}_{J+1},..., \widehat{\bm{\gamma}}_{S})$. 
By the assumption that $\frac{\rho_{\max} }{c_{\max}\lambda_{1}(1-\nu)} \to 0$, we have the first statement of Theorem~\ref{thm:2}
\begin{equation*}
  P(\forall g \in B_{2}, |2(\bm{y} - \bm{\Psi}\widehat{\bm{\gamma}})^{\top}\bm{\Psi}_{glq} - \lambda_{2}\widehat{\bm{\gamma}}_{glq}| < \lambda_{1}\nu c_{g}||\gammaestG||^{\nu-1}_{1}sgn(\gammaest_{glq})) \to 1
\end{equation*}
which is shown by the implied conditions as follows:
\begin{equation*}
P(\forall g \in B_{2}, ||\widehat{\bm{\gamma}}_{g}||^{2-\nu}_{2} > 0) \to 0
\end{equation*}

Now, we proceed to prove the asymptotic normality of the estimator $\widehat{\bm{\gamma}}_{B_{1}}$. First, we introduce the following notation:
\begin{equation*}
  \bm{r} \equiv \nu \times sgn(\widehat{\bm{\gamma}}_{B_{1}}) \odot ||\widehat{\bm{\gamma}}_{B_{1}}||_{1}^{\nu - 1}
\end{equation*}
to denote the gradient vector of the objective function with respect to the nonsparse coefficient $\widehat{\bm{\gamma}}_{B_{1}}$. The KKT conditions imply that
\begin{equation*}
 \bm{\Psi}_{B_{1}}^{\top}(\bm{y} - \bm{\Psi}_{B_{1}}\widehat{\bm{\gamma}}) - \lambda_{2}\bm{R}_{B_{1}}\widehat{\bm{\gamma}}_{B_{1}} = \lambda_{1}\bm{r}_{B_{1}},
\end{equation*}
and we can substitute $\bm{y}$ with the true data-generating model $\bm{y} = \bm{\Psi}\gammatrue + \bm{\epsilon} = \bm{\Psi}_{B_{1}}\bm{\gamma}^{\star}_{B_{1}} + \bm{\epsilon}$. Here, we introduce the shorter version of true coefficients and design matrix with  $\bm{\gamma}^{\star}_{B_{1}}$ and $\bm{\Psi}_{B_{1}}$ respectively. Furthermore, we use adopt our earlier notation $\bm{\Delta} = \widehat{\bm{\gamma}}_{B_{1}} - \bm{\gamma}^{\star}_{B_{1}}$ to denote the difference between the estimated and true coefficients in the significant set $B_{1}$. Then, we can rearrange the above equation
to obtain
\begin{equation*}
 \bm{\Psi}_{B_{1}}^{\top}(\bm{\Psi}_{B_{1}}\bm{\gamma}_{B_{1}}^{\star} + \bm{\epsilon} - \bm{\Psi}_{B_{1}}(\bm{\Delta} + \bm{\gamma}_{B_{1}}^{\star})) - \lambda_{2}\bm{R}_{B_{1}}(\bm{\Delta} + \bm{\gamma}_{B_{1}}^{\star}) = \lambda_{1}\bm{r}_{B_{1}},
\end{equation*}
which can be rearranged to
\begin{align*}
  \bm{\Psi}_{B_{1}}^{\top}(\bm{\epsilon} - \bm{\Psi}_{B_{1}}\bm{\Delta}) - \lambda_{2}\bm{R}(\bm{\Delta} + \bm{\gamma}_{B_{1}}^{\star}) &= \lambda_{1}\bm{r} \\
   \frac{1}{nM}(\bm{\Psi}_{B_{1}}^{\top}\bm{\Psi}_{B_{1}} + \lambda_{2}\bm{R}_{B_{1}})\bm{\Delta} &=  \frac{1}{nM}\bm{\Psi}_{B_{1}}^{\top}\bm{\epsilon} -  \frac{1}{nM}\lambda_{2}\bm{R}_{B_{1}}\bm{\gamma}_{B_{1}}^{\star} -  \frac{1}{nM}\lambda_{1}\bm{r}_{B_{1}}\\
  \sqrt{nM}\bm{\Delta} &= (\frac{1}{nM}\bm{\Psi}^{\top}\bm{\Psi} + \frac{\lambda_{2}}{nM}\bm{R}_{B_{1}})^{-1}\frac{1}{nM}\bm{\Psi}_{B_{1}}^{\top}\bm{\epsilon} + o(1) \\
  \sqrt{nM}\bm{\Delta} &= (\bm{\Psi}_{B_{1}}^{\top}\bm{\Psi}_{B_{1}} + \lambda_{2}\bm{R}_{B_{1}})^{-1}\bm{\Psi}_{B_{1}}^{\top}\bm{\epsilon} + o(1)\\
  \sqrt{nM}(\widehat{\bm{\gamma}}_{B_{1}} - \bm{\gamma}^{\star}_{B_{1}}) &\xrightarrow{d} \mathcal{N}(0, \sigma^{2}\bm{\Sigma}_{B_{1}}^{-1})
\end{align*}
which completes the proof of Theorem~\ref{thm:2}.
\end{proof}

\noindent\textit{\textbf{Proof of Theorem~\ref{thm:3}}}: The proof of Theorem~\ref{thm:3} is similar to that of Theorem~\ref{thm:2}, and the only difference is to fix $g$ to be in $B_{1}$ and repeat the proof over $l$. Hence, the proof is omitted here.

\section*{Acknowledgment}

The authors thank an associate editor and reviewers for insightful comments
and suggestions. This research is supported by the Natural Sciences and Engineering Research Council (NSERC) Discovery Grant (RGPIN-2024-03815, RGPIN-2022-05140) for H. Wang and T. Guan respectively. H.~L.~Shang is grateful for the financial support provided by the Australian Research Council Discovery Project DP230102250 and Future Fellowship FT240100338.

\begingroup
\setlength{\bibsep}{0pt plus 0.3ex} 
\setstretch{1.0} 
\bibliographystyle{elsarticle-harv}
\bibliography{IAMFPD}
\endgroup

\end{document}